\documentclass[sigconf]{acmart}

\makeatletter
\def\@ACM@checkaffil{
    \if@ACM@instpresent\else
    \ClassWarningNoLine{\@classname}{No institution present for an affiliation}%
    \fi
    \if@ACM@citypresent\else
    \ClassWarningNoLine{\@classname}{No city present for an affiliation}%
    \fi
    \if@ACM@countrypresent\else
        \ClassWarningNoLine{\@classname}{No country present for an affiliation}%
    \fi
}
\makeatother

\AtBeginDocument{%
  \providecommand\BibTeX{{%
    \normalfont B\kern-0.5em{\scshape i\kern-0.25em b}\kern-0.8em\TeX}}}

\copyrightyear{2023}
\acmYear{2023}
\setcopyright{acmlicensed}
\acmConference[KDD '23] {Proceedings of the 29th ACM SIGKDD Conference on Knowledge Discovery and Data Mining}{August 6--10, 2023}{Long Beach, CA, USA.}
\acmBooktitle{Proceedings of the 29th ACM SIGKDD Conference on Knowledge Discovery and Data Mining (KDD '23), August 6--10, 2023, Long Beach, CA, USA}
\acmPrice{15.00}
\acmISBN{979-8-4007-0103-0/23/08}
\acmDOI{10.1145/3580305.3599238}
\usepackage{balance}
\usepackage{multirow}
\usepackage{enumitem}
\usepackage{graphicx}  
\usepackage{float}  
\usepackage{subfigure}  
\usepackage{stfloats}
\usepackage{url}
\usepackage{multirow}
\usepackage{booktabs}
\usepackage[ruled,linesnumbered]{algorithm2e} 

\begin{document}




\title{3D-IDS: Doubly Disentangled Dynamic Intrusion Detection}

\author{Chenyang Qiu}
\affiliation{%
  \institution{Beijing University of Posts and Telecommunications}}
\email{cyqiu@bupt.edu.cn}
 
\author{Yingsheng Geng}
\affiliation{%
  \institution{Beijing University of Posts and Telecommunications}}
\email{gyswasdfre2255@bupt.edu.cn}

\author{Junrui Lu}
\affiliation{%
  \institution{Beijing University of Posts and Telecommunications}}
\email{672@bupt.edu.cn}

\author{Kaida Chen}
\affiliation{%
  \institution{Beijing University of Posts and Telecommunications}}
\email{chenkaida@bupt.edu.cn}

\author{Shitong Zhu}
\affiliation{%
  \institution{Beijing University of Posts and Telecommunications}}
\email{1337612789@bupt.edu.cn}

\author{Ya Su}
\affiliation{%
  \institution{HUAWEI Technologies Co., Ltd.}}
\email{suya9@huawei.com}

\author{Guoshun Nan} 
\authornote{Guoshun Nan is the corresponding author.}
\affiliation{%
  \institution{Beijing University of Posts and Telecommunications}}
\email{nanguo2021@bupt.edu.cn}

\author{Can Zhang}
\affiliation{%
  \institution{Beijing University of Posts and Telecommunications}}
\email{zhangcan_bupt@bupt.edu.cn}

\author{Junsong Fu}
\affiliation{%
  \institution{Beijing University of Posts and Telecommunications}}
\email{fujs@bupt.edu.cn}

\author{Qimei Cui}
\affiliation{%
  \institution{Beijing University of Posts and Telecommunications}}
\email{cuiqimei@bupt.edu.cn}

\author{Xiaofeng Tao}
\affiliation{%
  \institution{Beijing University of Posts and Telecommunications}}
\email{taoxf@bupt.edu.cn}
\renewcommand{\shortauthors}{Chenyang Qiu et al.}

\begin{abstract}
Network-based intrusion detection system (NIDS) monitors network traffic for malicious activities, forming the frontline defense against increasing attacks over information infrastructures. Although promising, our quantitative analysis shows that existing methods perform inconsistently in declaring various unknown attacks (e.g., $9\%$ and $35\%$ F1 respectively for two distinct unknown threats for an SVM-based method) or detecting diverse known attacks (e.g., $31$\% F1 for the Backdoor and $93$\% F1 for DDoS by a GCN-based state-of-the-art method), and reveals that the underlying cause is \textit{entangled distributions of flow features}. This motivates us to propose 3D-IDS, a novel method that aims to tackle the above issues through two-step feature disentanglements and a dynamic graph diffusion scheme. Specifically, we first disentangle traffic features by a non-parameterized optimization based on mutual information, automatically differentiating tens and hundreds of complex \textbf{features of various attacks}. Such differentiated features will be fed into a memory model to generate  representations, which are further disentangled to highlight the \textbf{attack-specific features}. Finally, we use a novel graph diffusion method that dynamically fuses the \textbf{network topology} for spatial-temporal aggregation in evolving data streams. By doing so, we can effectively identify various attacks in encrypted traffics, including unknown threats and known ones that are not easily detected. Experiments show the superiority of our 3D-IDS. We also demonstrate that our two-step feature disentanglements benefit the explainability of NIDS.

\end{abstract}

\ccsdesc[300]{Security and privacy~Mobile and wireless security}


\keywords{Intrusion Detection; Anomaly Detection; Network Security}
\maketitle
\section{Introduction}
\label{sec:intro}

\begin{figure}
  \centering
  \setlength{\abovecaptionskip}{5pt}  
  \setlength{\belowcaptionskip}{0pt}
  \includegraphics[scale=0.41]{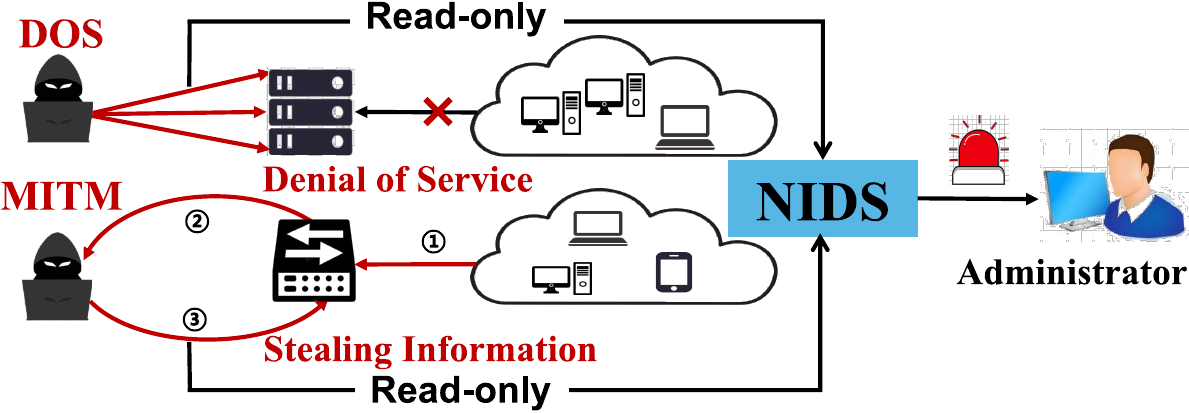}\\
  \caption{Illustration of two network attacks DoS and MITM. DoS floods the target with massive traffic to overwhelm an online service, and MITM eavesdrops on the communication between two targets and steals private information. An NIDS can be easily deployed in a single location to collect statistical
features and alert the administrator for potential threats.}\label{fig:intro-illustrate}
  \vspace{-10pt}
\end{figure}

Network attacks \cite{bhushan2018recent}, such as password cracking \cite{kelley2012guess}, man-in-the-middle attacks (MITM) \cite{cekerevac2017internet}, and denial of service (Dos) \cite{prasad2014and}, refer to unauthorized attempts on various digital assets in an organization's networks, to steal data or perform malicious actions. Network attacks can also be broadly regarded as network anomalies \cite{li2006detection}, as they may involve different characteristics from most other traffics. It was reported that $31\%$ of companies worldwide are now being attacked at least one time a day \cite{wueest2014targeted}, due to the growing trends of mobile online business. This urgently calls for an intelligent system that can help administrators automatically filter out these network anomalies in a huge amount of online traffics. A network-based intrusion detection system (NIDS) \cite{kumar2016machine}, which monitors network traffic and identifies malicious activities, facilitates administrators to form the frontline defense against increasing attacks over information infrastructures (e.g., sensors and servers). Hence, NIDS is widely applied in many information systems of governments and e-commercial business sectors \cite{zhao2009unknown}. Figure \ref{fig:intro-illustrate} illustrates that two network attacks threaten information systems, and a NIDS builds a frontline defense against these threats.

\begin{figure*}
    \centering
    \subfigbottomskip=0pt    
    \subfigcapskip=-3pt 
    \setlength{\abovecaptionskip}{8pt}  
    \setlength{\belowcaptionskip}{0pt}
    \subfigure[F1-Score]{
    \includegraphics[height=1.8cm,width=0.32\linewidth]{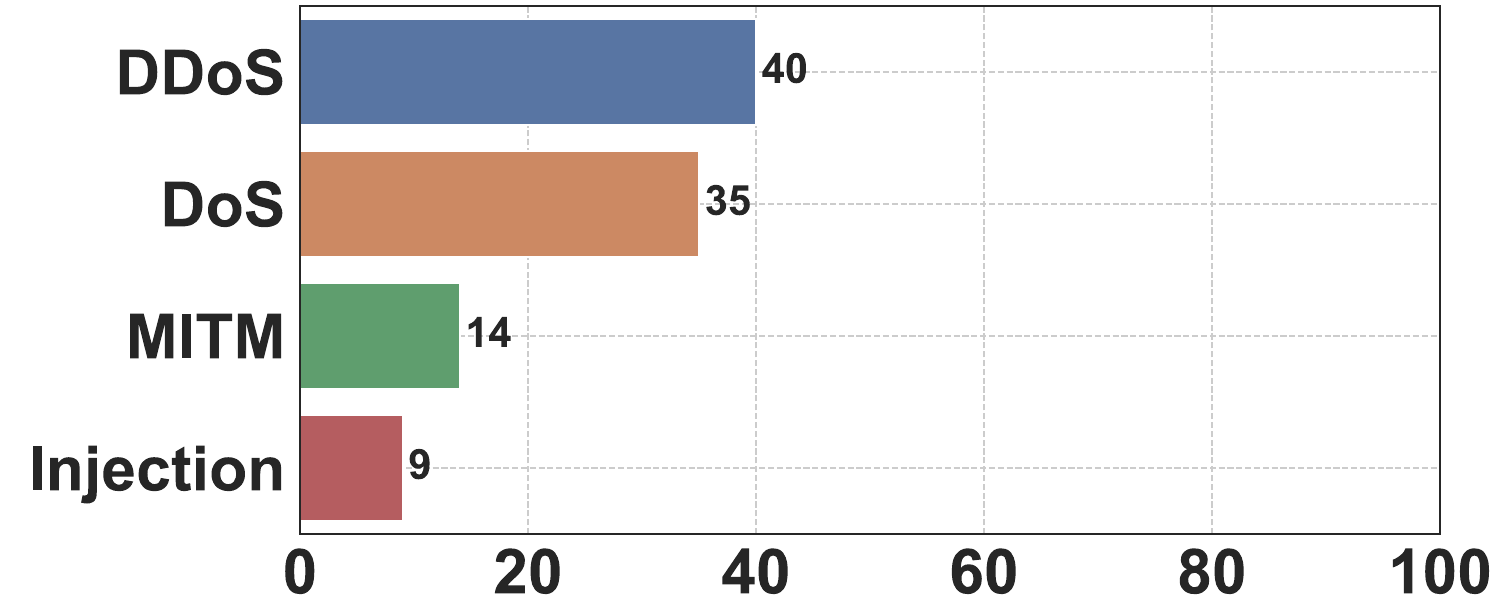}
    }
     \subfigure[Normalized feature distribution of MITM attacks]{
     \includegraphics[height=1.8cm,width=0.32\linewidth]{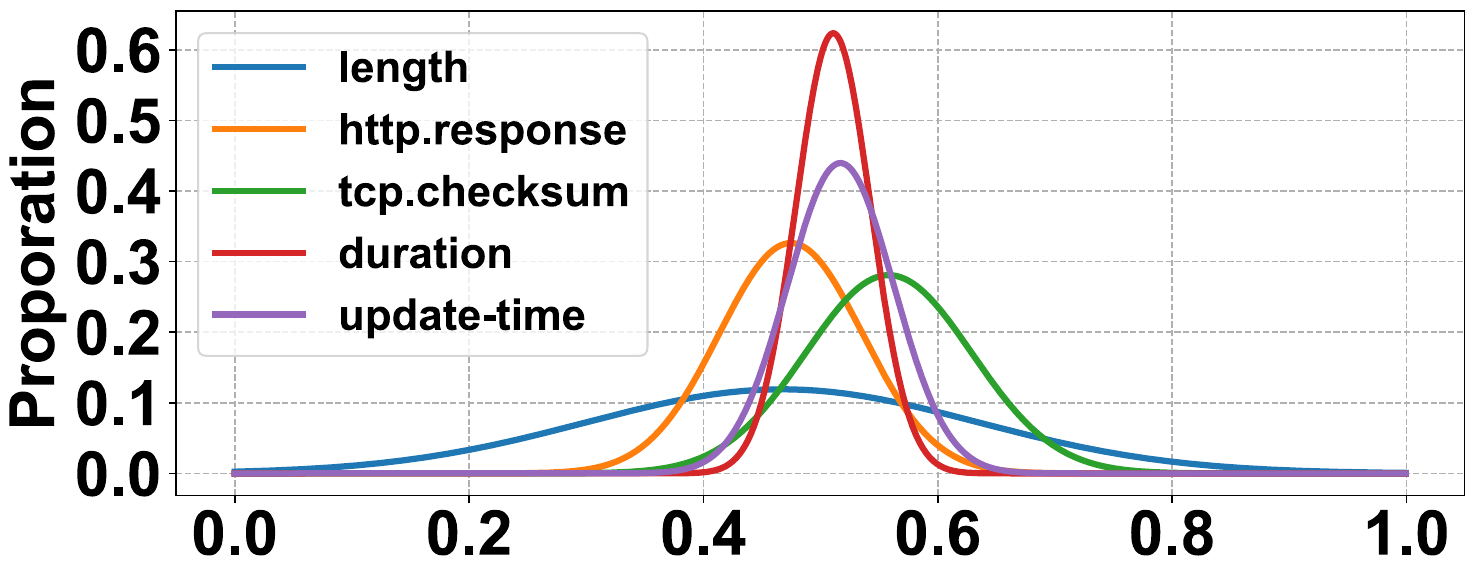}
    }
    \subfigure[Normalized feature distributions of DDoS attacks]{
    \includegraphics[height=1.8cm,width=0.32\linewidth]{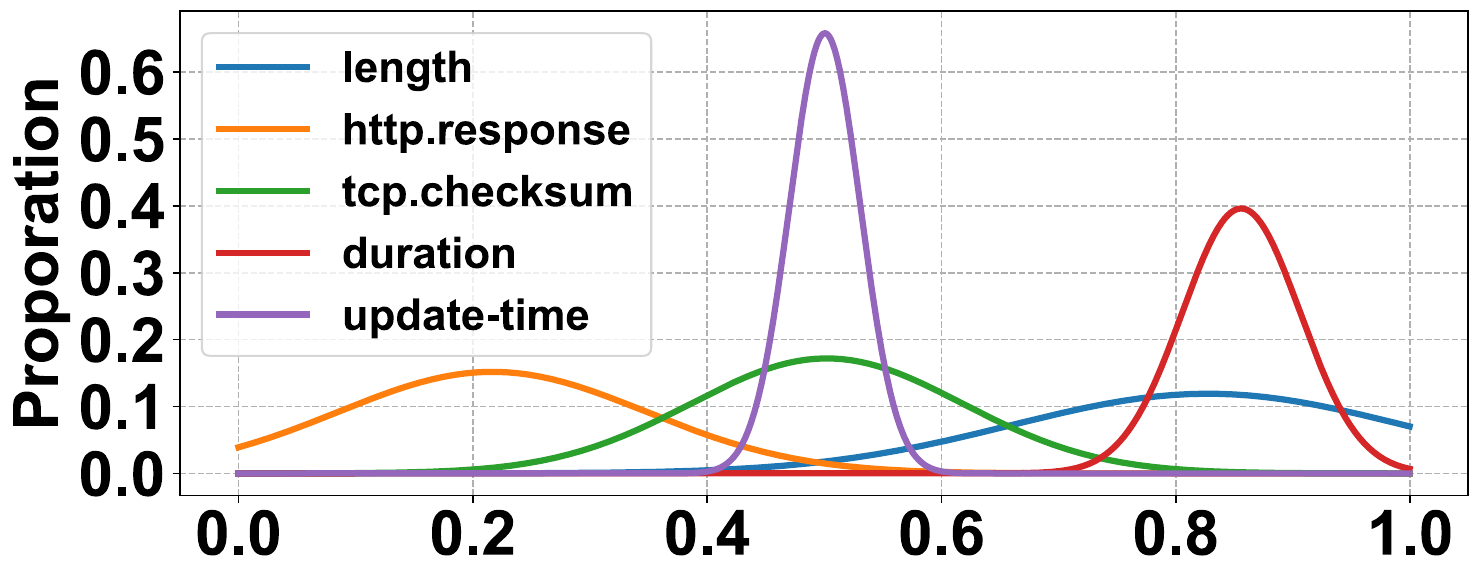}
    }
    \subfigure[F1-Score]{
    \includegraphics[height=1.8cm,width=0.32\linewidth]{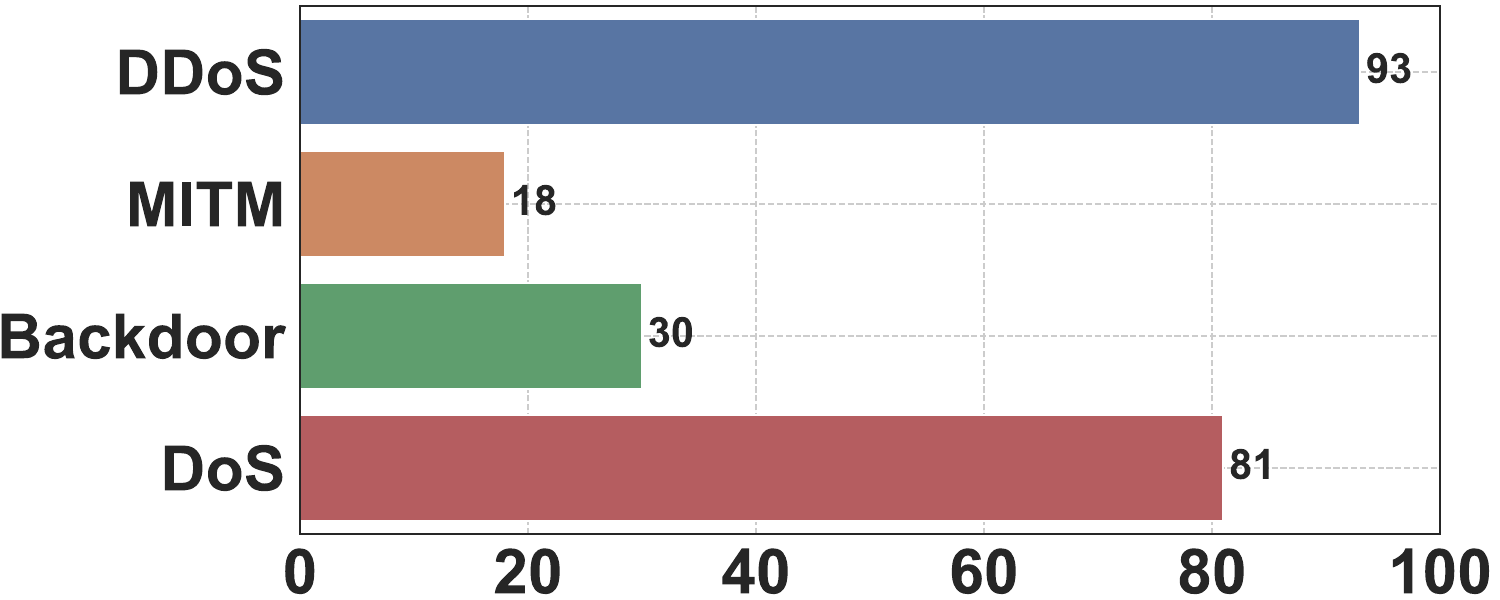}
    }
    \subfigure[Representation correlation map of MITM attacks]{
    \includegraphics[height=1.8cm,width=0.32\linewidth]{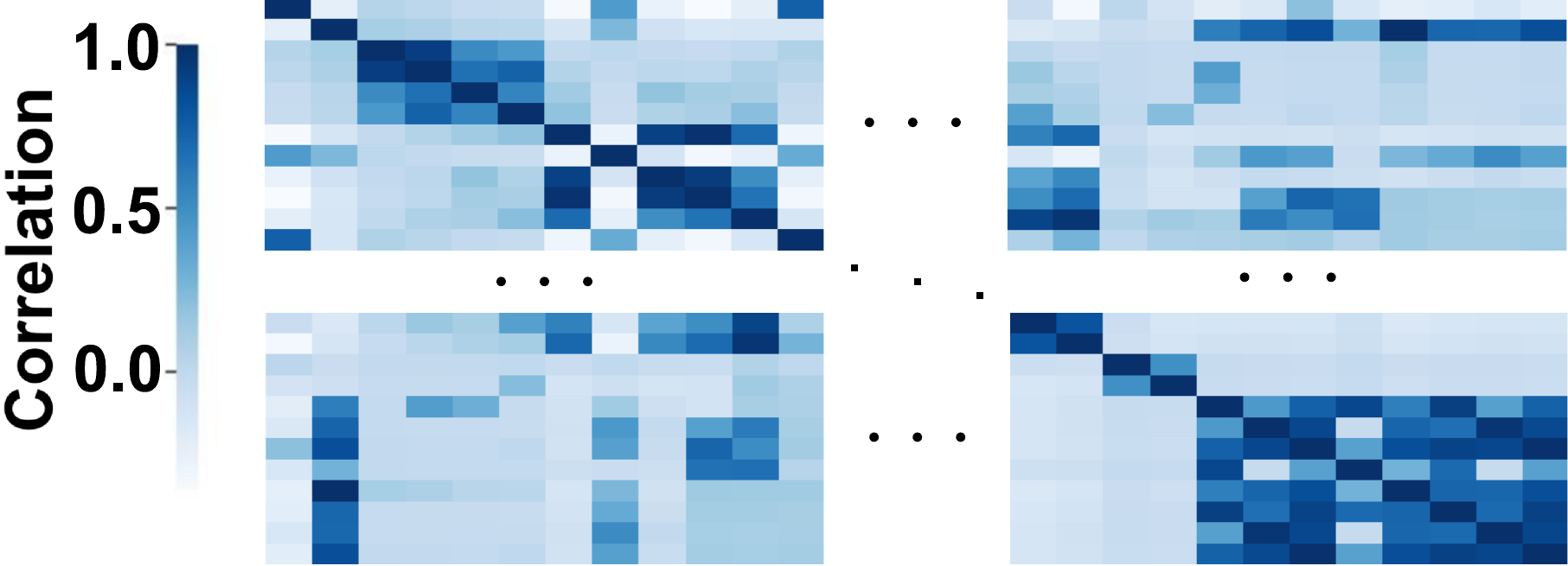}
    }
    \subfigure[Representation correlation map of DDoS attacks] {
    \includegraphics[height=1.8cm,width=0.32\linewidth]{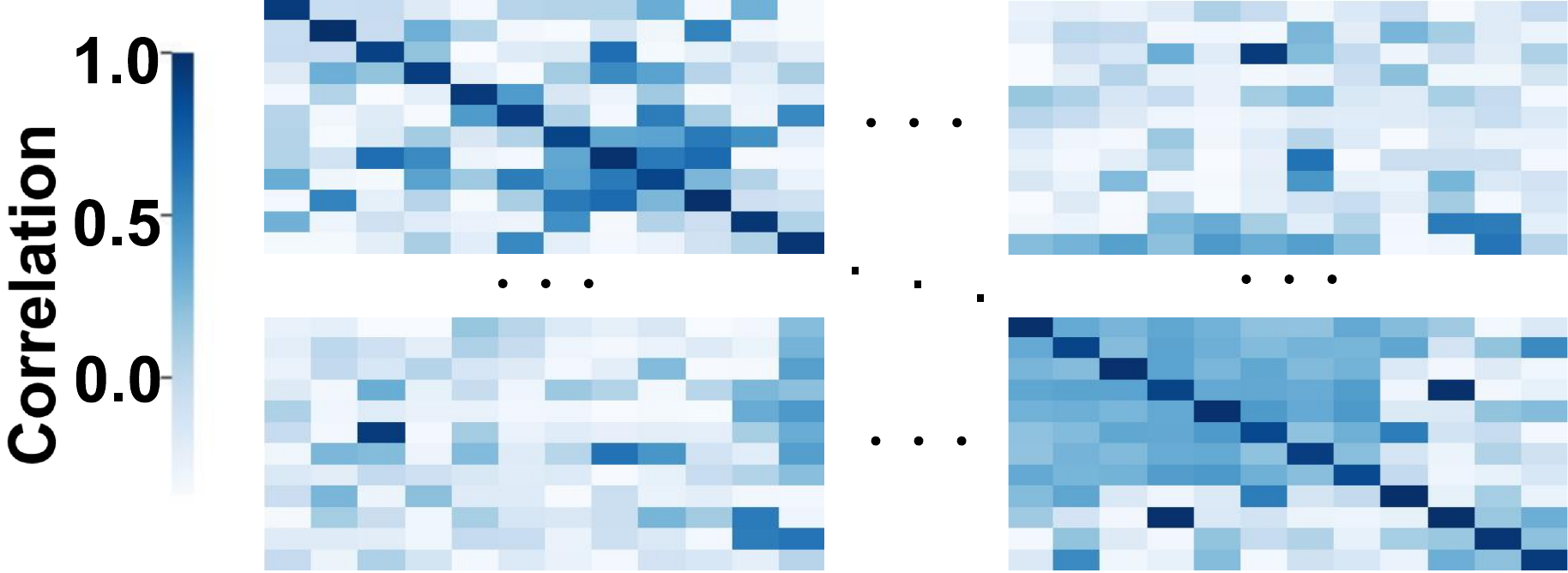}
    }
    \\
    \caption{Quantitative analysis on CTC-TON-IOT.
    (a) Comparisons of detecting various  attacks, which are regarded as an unknown type in evaluation. Specifically, we train an SVM model without using the data points of these attacks, and evaluate the instances of these attacks on the test set. (b) and (c) show the feature distributions of two attacks, MITM and DDoS, respectively. (d) Comparisons of detecting various known attacks on the previous state-of-the-art deep learning model E-GraphSAGE, (e) and (f) are correlation maps of representations of the two attacks, where the representations are generated by E-GraphSAGE.     
    }
    \label{fig:intro-analysis}
\end{figure*}

Existing NIDS can be categorized into two types, i.e., signature-based ones \cite{ioulianou2018signature,erlacher2018fixids,masdari2020survey} and anomaly-based ones \cite{samrin2017review,van2017anomaly,aljawarneh2018anomaly,eskandari2020passban}. The former detects network attacks based on pre-defined patterns or known malicious sequences stored in a database, such as the number of bytes in traffic. These patterns in the NIDS are referred to as signatures. The latter anomaly-based NIDS learns to track attacks with machine learning techniques. Early statistical \cite{cekerevac2017internet,prasad2014and} approaches, such as Support Vector Machine (SVM), Logistic Regression (LR), and Decision Tree (DT), rely on carefully designed handcrafted features to learn classification boundaries. Recent deep learning-based methods \cite{ahmad2021network,Bhatia_2021} use millions of neural parameters to mine the knowledge underlying the training samples, and have achieved great success in automatically modeling complex correlations for tens and thousands of features. The state-of-the-art E-GraphSAGE \cite{lo2022graphsage} employs graph convolution networks (GCNs) to learn the feature representations for better prediction.

Although promising, we observe that existing anomaly-based approaches yield inconsistent results in identifying distinct attacks. For statistical methods, Figure \ref{fig:intro-analysis} (a) demonstrates that the
detection performance of an SVM-based method \cite{ioannou2021network} for an unknown attack can be as low as $9\%$ in terms of F1 on CTC-ToN-IOT  \cite{moustafa2021new}, a popular benchmark for NIDS. While the model achieves $40\%$ F1 in declaring another unknown threat (DDoS) on the same benchmark. Regarding the deep learning-based methods, Figure \ref{fig:intro-analysis} (d) demonstrates that E-GraphSAGE achieves a lower than $20$\% F1 score for MITM attacks, and a higher than $90$\% F1 score for DDoS on the CTC-ToN-IOT dataset.




To investigate the underlying cause of why existing methods perform inconsistently for various attacks, including unknown ones and known ones, we depict feature distributions and visualize the representations. Figure \ref{fig:intro-analysis} (b) and (c) depict statistical distributions of the two unknown attacks for the SVM-based model during testing. We observe that feature distributions of MITM attacks are entangled, while the ones of DDoS are more separated. It can be inferred that statistical distributions of traffic features are one of the main underlying causes of performance variations. Separated distributions benefit the unknown attack identification, while entangled ones are indistinguishable and unable to help the NIDS to make accurate decisions. We refer to such a phenomenon as \textbf{the entangled distribution of statistical features}. To analyze the reason for performance variations of acknowledged attacks during testing, we use Pearson correlation heat map \cite{cohen2009pearson} to visualize the representations of MITM and DDoS respectively, where the representations are generated by the encoder of E-GraphSAGE. Figure \ref{fig:intro-analysis} (e) and Figure \ref{fig:intro-analysis} (f) demonstrate the two correlation maps. Interestingly, we observe that the coefficients of MITM representations are much larger than those of DDoS. We further compare MITM with other attacks, including Backdoor and Dos, and find those high coefficients in the representation will lead to lower intrusion detection scores. We refer to such a phenomenon as \textbf{the entangled distribution of representational features}, which can be considered as another main cause for the degradation of attack classification.



In light of the above discussion, a critical question arises: "\textit{How can an intrusion detection model automatically address the above issue, i.e., two entangled distributions, to benefit the detection of both unknown and known attacks}". Achieving this goal is challenging. To mitigate the issue of the first entangled distribution, we need to differentiate tens and thousands of features involved in real-time network traffic, without prior knowledge of the statistical distributions. 
Such a problem is largely under-explored in the field of NIDS. For the second entangled distribution, there are some remotely related methods \cite{liu2019bidirectional,brodbeck2020continuous,sharma2020comprehensive,guo2019learning} in other fields, including computer vision and natural language processing. However, these approaches mainly focus on object-level representation learning, and hence they are hardly directly applied to intrusion detection to tackle this challenging problem. 


We answer the above question and address the two challenging problems with a novel method called 3D-IDS (\textbf{D}oubly \textbf{D}isentangle \textbf{D}ynamic IDS). Specifically, we first disentangle the statistical flow features with a non-parametric optimization based on mutual information, so that entangled distributions can be automatically separated for modeling the general features of various attacks. We refer to this step as statistical disentanglement. Then we further learn to differentiate the  representations by a loss function, aiming to highlight the salient features for specific attacks with smaller coefficients. We refer to this step as representational disentanglement. We finally introduce a novel graph diffusion method that fuses the graph topology for spatial-temporal aggregation in evolving traffic. Extensive experiments on five benchmarks show the superiority of our 3D-IDS. The main contributions of this paper are as follows.
\begin{itemize}
    \item We propose 3D-IDS, a novel method that aims to mitigate the entangled distributions of flow features for NIDS. To the best of our knowledge, we are the first to quantitatively analyze such an interesting problem and empirically reveal the underlying cause in the field of intrusion detection.  
    \item We present a double-feature disentanglement scheme for modeling the general features of various attacks and highlighting the attack-specific features, respectively. We additionally introduce a novel graph diffusion method for better feature aggregation.
    \item We conduct extensive quantitative and qualitative experiments on various benchmarks and provide some helpful insights for effective NIDS. 
\end{itemize}

\
\vspace{-15pt}
\section{RELATED WORK}
\subsection{Network Intrusion Detection System}
Existing NIDS can be classified into two groups, i.e., signature-based ones \cite{ioulianou2018signature,erlacher2018fixids,masdari2020survey} and anomaly-based ones \cite{samrin2017review,van2017anomaly,aljawarneh2018anomaly,eskandari2020passban}. The latter involves statistical methods \cite{cekerevac2017internet,prasad2014and} and deep learning-based ones\cite{ahmad2021network,Bhatia_2021}. Early deep learning studies model the traffic as independent sequences \cite{ma2020deep,hu2021multiple,li2022graphDDoS,doriguzzi2020lucid}. Recent popular studies rely on GCN to aggregate traffic information \cite{kalander2020spatio,cong2021dynamic,song2020spatial}. Most related to our work is Euler \cite{kingeuler}, which builds a series of static graphs based on traffic flow and then performs information aggregation. Our 3D-IDS differs from the above methods in two aspects: 1) We build dynamic graphs rather than static ones, and such a dynamic aggregation can capture fine-grained traffic features for attack detection. Furthermore, we fuse the network layer information into our graph aggregation. 2) We also introduce a two-step disentanglement scheme, including statistical disentanglement and representational one, to benefit the detection of both known and unknown attacks.  

\subsection{Disentangled Representation Learning}
Disentanglement aims to learn representations that separate the underlying explanatory factors responsible for variation in the data. Previous studies \cite{bengio2013representation,chen2020weakly,gilpin2018explaining,gondal2020transfer,locatello2020weakly,lipton2018mythos,creager2019flexibly} focus on the generative models by employing constraints on the loss functions, such as $\beta$-VAE \cite{higgins2017beta}. Some recent approaches \cite{suter2019robustly,ridgeway2018learning,trauble2021disentangled,zhou2022link} have studied disentangled GNNs to capture the intrinsic factors in graph-structured data. Most related to our work is DisenLink 
~\cite{zhou2022link}, which disentangled the original features into a fixed number of factors, with selective factor-wise message passing for better node representations. While our 3D-IDS uses a double disentanglement method, which first disentangles the statistical features via non-parametric optimization, and then learns to highlight the attack-specific features with a regularization.

\subsection{Dynamic Graph Convolution Networks}
Dynamic graph convolution networks (GCNs) focus on evolving graph streams. There is a line of early studies in GCNs on dynamic graphs, which incorporate temporal information into graphs. These methods can be categorized into the spatio-temporal decoupled ones~\cite{pareja2020evolvegcn,kalander2020spatial,cong2021dynamic} and the spatio-temporal coupled ones~\cite{song2020spatial,poli2019graph,gu2020implicit,chen2022optimization,chamberlain2021grand}. The former employs two separate modules to capture temporal and spatial information respectively, and the latter ones incorporate spatial-temporal dependencies by proposing a synchronous modeling mechanism. Our 3D-IDS is mainly inspired by the GIND \cite{chen2022optimization}, which adaptively aggregates information via a non-linear diffusion method. The key difference between our GCN approach and GIND is: we introduce the non-linear graph diffusion method into a multi-layer graph that considers the network topology for dynamic intrusion detection.

\section{PRELIMINARIES}
\subsection{Multi-Layer Graphs}

We consider a single-layer network modeled by a graph  $\mathrm{G}=(\mathrm{V}, \mathrm{E}, \omega)$, where  $V$  is the set of nodes and  $E \subset \mathrm{V} \times \mathrm{V}$  is the set of edges. Here $\omega: \mathrm{V} \times \mathrm{V} \mapsto \mathrm{R}$  is an edge weight function such that each edge  $e_{uv} \in \mathrm{E}$ has a weight $ \omega_{uv}$. 
A multi-layer network can be modeled by a multi-layer graph:

\begin{equation}
    \mathbb{A}=\left(\begin{array}{ccccc}
    A_{(1,1)} & \cdots & A_{(1, k)} & \cdots & A_{(1, m)} \\
    \vdots & \ddots & \vdots & \ddots & \vdots \\
    A_{(l, 1)} & \cdots & A_{(k, k)} & \cdots & A_{(l, m)} \\
    \vdots & \ddots & \vdots & \ddots & \vdots \\
    A_{(m, 1)} & \cdots & A_{(m, k)} & \cdots & A_{(m, m)}
    \end{array}\right).
\end{equation}
where $A_{i,i}$ refers to the intra-layer adjacency matrix and  $A_{i,j} (i\neq j)$ refers to cross-layer adjacency matrix.  

 \subsection{Edge Construction}

We have constructed a multi-layer dynamic graph by defining the devices as nodes and the communications between two devices as edges.
Specifically, we first transform the Netflow data to edges by the following definition:  
\begin{equation}
\mathbf{E}_{i j}(t)=(v_{i}, l_{i}, v_{j}, l_{j}, t, \Delta t, \mathbf{F}_{ij}(t)).
\label{eqa}
\end{equation}
First, we concatenate the source IP and source port in the original traffic flows as the source identity for the device $i$. Similarly, we can obtain the destination identity for the device $j$. We denote $v_i$ and $v_j$ as the source and destination nodes respectively. Secondly, $l_i$ denotes the layer of device $i$, and $l_i=0$ indicates that $i$ is a terminal device such as a PC, a server, or an IoT device. Here $l_i=1$ indicates that $i$ is an intermediate device such as a router in the communication link. Specifically, we assign devices with the router address $192.168.0.1$ or with many stable connections in layer $1$. Then $t$ refers to the timestamp of traffic and $\Delta t$ indicates the traffic duration time. Finally, $\mathbf{F}_{ij}(t)$ is the traffic features.

\subsection{Problem Formulation}
We first define the devices as nodes and the communications with timestamps between any pair of devices as edges. We use $\mathrm{T}$ to represent the maximum timestamp. An Edge sequence $\mathbb{E}$  is denoted by  $\left\{\mathcal{E}^{t}\right\}_{t=1}^{T}$, where each  $\mathcal{E}^{t}$ represents a network traffic. 
Also, after each edge, there is a corresponding multi-layer graph, then the corresponding multi-layer graph stream  $\mathbb{G}$ takes the form of  $\left\{\mathcal{G}^{t}\right\}_{t=1}^{T} $, where each  $\mathcal{G}^{t}=\left(\mathcal{V}^{t}, \mathcal{E}^{t}\right)$  represents the multi-layer graph at timestamp $t$. 
A multi-layer adjacency matrix  $\mathbb{A}^{t} \in \mathbb{R}^{m \times n} $ represents the edges in  $\mathcal{E}^{t}$, where  $\forall(i, j, w) \in \mathcal{E}^{t}, \mathbb{A}^{t}[i][j]=w_{ij}$ and $w_{ij}$ is the weight of the matrix. The goal of intrusion detection is to learn to predict the edge  $\mathcal{E}^{t}$ as a benign traffic or an attack in binary classification, and a specific type under the multi-classification.  



\section{MODEL}
\begin{figure*}[!htb]
 \centering
     \includegraphics[width=\linewidth,height=0.40\textwidth]{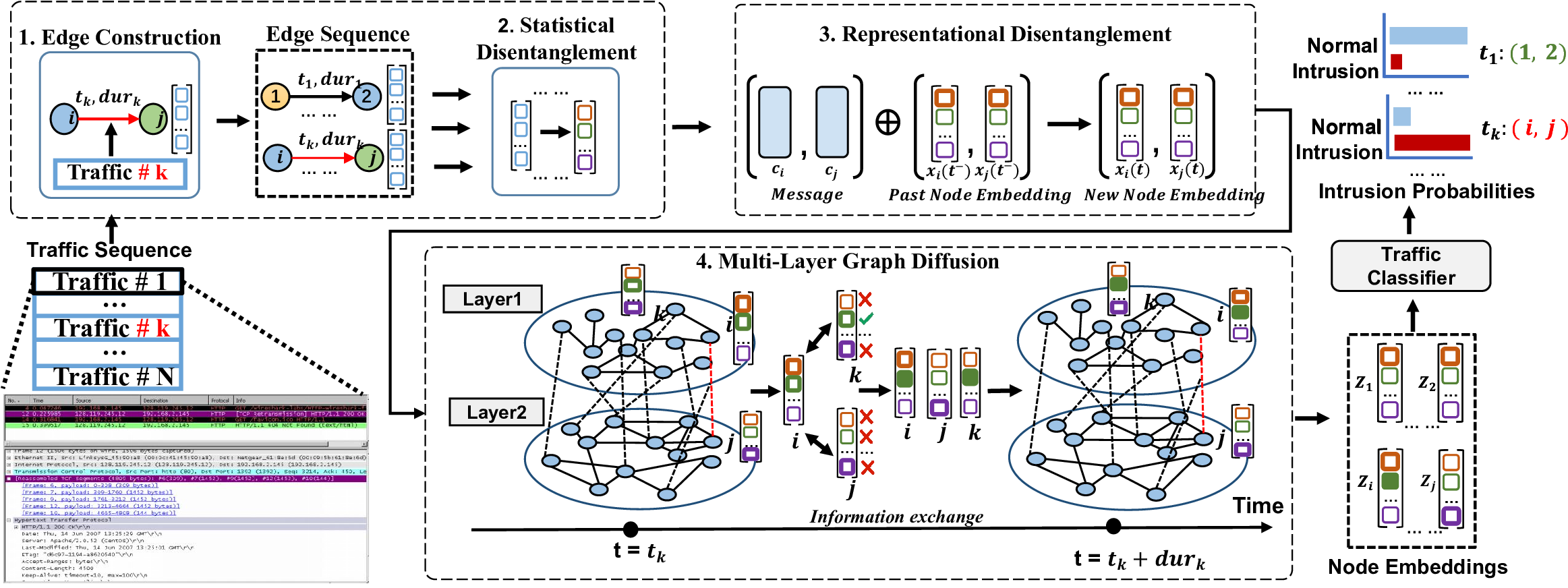}
      \vspace{-10pt}
     \caption{Overview of the proposed 3D-IDS, which consists of five modules. 1) Edge construction module builds edges based on traffic flow. 2) Statistical disentanglement module differentiates values in vectors to facilitate the identification of various attacks. 3) Representational disentanglement module learns to highlight attack-specific features. 4) Multi-Layer graph diffusion module fuses the network topology for better aggregation over evolving dynamic traffic. 5) Traffic classifier takes the traffic representation as an input to yield the detection results. }
     \label{model}
 \end{figure*}

In this section, we present the 3D-IDS, which consists of four main modules. Figure \ref{model} shows the architecture of our proposed model. Next, we dive into the details of these modules.

\subsection{Statistical Disentanglement}
\label{Edge Representation Disentangle module}
 

%
As we discussed in Section \ref{sec:intro}, statistical distributions of traffic features are one of the main underlying causes of performance variations. i.e., separated distributions benefit the unknown attack identification, while entangled ones are indistinguishable and thus unable to help the NIDS to make accurate decisions. Therefore, our aim is there to disentangle the traffic features and make them distinguishable.


To separate the features of traffic without any prior knowledge, we formulate the differentiation as a constrained non-parametric optimization problem and approximate the optimal results by solving the Satisfiability Modulo Theory (SMT) \cite{de2008z3}. We perform a min-max normalization on the edge feature $\mathbf{F}_{ij}(t)$. For convenience, we denote the normalized edge feature as $\mathcal{F}$, and $\mathcal{F}_i$ is the $i$-th normalized element.




We need a weight matrix $\mathbf{w}$ to generate the disentangled representation of $\mathcal{F}$. Our key optimization objectives are to minimize the mutual information between the elements of traffic features and also bound the range of $\mathbf{w}$ when we perform aggregation. We start by constraining the weight matrix $\mathbf{w}$ with the range of the superposition function as follows:
\begin{gather}
    W_{\min } \leq \mathbf{w}_{i} \leq W_{\max } \quad(1 \leq i \leq N),\quad
    \sum_{i=1}^{N} \mathbf{w}_{i} \mathcal{F}_{i} \leq B,
\end{gather}
where $W_{\min }$, $W_{\max }$ and $B$ are constants, $N$ is the edge feature dimension.
We then constrain the order-preserving properties of generated representations:
\begin{equation}
  \mathbf{w}_{i} \mathcal{F}_{i} \leq \mathbf{w}_{i+1} \mathcal{F}_{i+1}\quad(1 \leq {i} \leq {N-1}).
\end{equation}

Finally, we maximize the distance between components in the vector $\mathbf{w}$, consequently minimizing the mutual information between each two feature elements. In this way, we can disentangle the distribution of element-wised features. The optimization objective can be expressed as follows:

\begin{equation}
\begin{split}
    \widetilde{\mathbf{w}}=\arg \max (\mathbf{w}_{N} \mathcal{F}_{N}-\mathbf{w}_{1} \mathcal{F}_{1}-
    \\\sum_{i=2}^{N-1} \left|2 \mathbf{w}_{i} \mathcal{F}_{i}-\mathbf{w}_{i+1} \mathcal{F}_{i+1}-\mathbf{w}_{i-1} \mathcal{F}_{i-1}\right|).
  \end{split}
  \label{target}
\end{equation}
We are unable to determine the convexity of the optimization
object due to its closed form. Therefore, we transform the above problem into an SMT problem with an optimization objective (\ref{eqb}) and a subjection (\ref{constrain}) to approximate the optimal results.

\begin{equation}
    \begin{split}
        \widetilde{\mathbf w}=\arg \max (\mathbf{w}_{N} \mathcal{F}_{N}-\mathbf{w}_{1} \mathcal{F}_{1}+
        \\ \sum_{i=2}^{N-1} 2 \mathbf{w}_{i} \mathcal{F}_{i }-\mathbf{w}_{i-1} \mathcal{F}_{i-1}-\mathbf{w}_{i+1} \mathcal{F}_{i+1}),
    \end{split}
    \label{eqb}
\end{equation}
subjects to:
\begin{equation}
  \left\{\begin{array}{lll}
    \mathbf{w}_{i} & \in & {\left[W_{\min }, W_{\max }\right]} \\
    \sum_{i=1}^{N} \mathbf{w}_{i} \mathbf{n}_{i} & \leq & B \\
    \mathbf{w}_{i} \mathcal{F}_{i} & \leq & \mathbf{w}_{i+1} \mathcal{F}_{(i+1)} \\
    2 \mathbf{w}_{i} \mathcal{F}_{i} & \leq & \mathbf{w}_{i-1} \mathcal{F}_{i-1}+\mathbf{w}_{i+1} \mathcal{F}_{i+1}.
  \end{array}\right.
  \label{constrain}
\end{equation}
For the above formation, we can generate the disentangled edge representation $\mathbf h_{i,j}$ by solving the above problem, which can be expressed as $\mathbf h_{i,j}=\mathbf{w}\odot\mathcal{F}$, where the symbol $\odot$ represents the Hadamard product. 

Equipped with the above non-parametric optimization, we can differentiate tens and hundreds of complex features of various attacks, mitigating the entangled distribution of statistical features. Such statistical disentangled features facilitate our model to be more sensitive to various attacks. 


\subsection{Representational Disentanglement}
So far we have constructed edges and statistically differentiated the features of traffic flows. This module generates contextualized \textbf{node representations} $\mathbf X$ from edge representations. This involves three steps:

\textbf{1) Generating updating messages:}
For an incoming traffic flow, we will build an edge or update the corresponding edge, which may lead to a dramatic change in the node representations involved in this interaction. We can update the node representations by utilizing this change. Therefore, we name the abrupt change an updating message and denote it as $c(t)$. Specifically, we generate $c(t)$ by incorporating historical memory and disentangled edge representation $\mathbf h_{i,j}$. The messages of node $i$ and $j$ are as follows:        
       \begin{align}
              \mathbf{c}_{i}(t)&=\operatorname{Msg}\left(\mathbf{m}_{i}\left(t^-\right),\mathbf{m}_{j}\left(t^-\right),t,\Delta t,l_i, l_j, \mathbf{h_{i,j}}\right),\\
              \mathbf{c}_{j}(t)&=\operatorname{Msg}\left(\mathbf{m}_{j}\left(t^-\right),\mathbf{m}_{i}\left(t^-\right),t,\Delta t,l_i, l_j,\mathbf{h_{i,j}}\right),
              \label{eqc}
        \end{align}
        where $\mathbf{h}_{i,j}$ is the disentangled edge representation. $\Delta t$ is the edge duration time, $l_i,l_j$ is the layer marks of edge, \(m_i(t^-) \), \(m_j(t^-) \)is the historical memory of the two interacting nodes, $\operatorname{Msg}$ is a learnable function, and we use RNN. The initial memory $\mathbf m(0)$ is set as $\mathbf 0$.
 
        
\textbf{2) Updating node memory:}
        We update the node memory by merging the latest message with the historical memory and then encode the merging information, which can be expressed as follows:
               \begin{equation}
          \mathbf{m}_{i}(t)=\operatorname{Mem}\left(\mathbf{c}_{i}(t),  \mathbf{m}_{i}\left(t^{-}\right)\right),
          \label{eqd}
        \end{equation}
        where $\operatorname{Mem}$ is an encoder, and here we use GRU. We also update the memory of node $j$ via (\ref{eqd}).
          
\textbf{3) Generating second disentangled node representations:}
We can generate the node representation by utilizing the updated memory in (\ref{eqd}), and the representation of node $i$ in time $t$ can be expressed as $\mathbf{x}_i(t)=\mathbf{x}_i(t^-)+\mathbf{m}_i(t)$, where
$\mathbf{{x}}_i(t^-)$ is the historical representation of node $i$. The representation of node $j$ can also be generated in a similar way. The initial value of node representations $\mathbf X (0)$ is $\mathbf 0$.

We aim to preserve the disentangled property in node representations. However, the update operations above may entangle them at the element level again. 
Therefore, we propose the second representational disentanglement, which aims to highlight the attack-specific features for the following end-to-end training. It also ensures that the element-wised representations are close to orthogonal.

\vspace{-8pt}
        \begin{equation} 
             \mathcal{L}_{\text {Dis }}=\frac{1}{2}\left\|\mathbf{X} (t){\mathbf{X}({t^-})}^{\top}-\mathbf{I}\right\|_{F}^{2}.
             \label{eqe}
        \end{equation}

The above regularization encourages the model to learn smaller coefficients between every two elements of node representations. Such representations can be more differentiated. As the module is supervised by the signal of a specific attack, it can learn to highlight the attack-specific features, so as to improve the detection scores, as discussed in Section \ref{sec:intro}. By doing so, we are able to mitigate the entangled distribution of representational features as mentioned at the beginning.   

\subsection{Multi-Layer Graph Diffusion}
\label{Multi-Layer Graph Diffusion module}
So far, we have generated the disentangled node representations with temporal information. For further fusing the multi-layer topological structure information, we propose a multi-layer graph diffusion module, and please note that we still preserve the disentangled property by customized designs. 

We utilize the following graph diffusion method to fuse the topological information in evolving graph streams, which can capture the fine-grained continuous spatio-temporal information. The previous dynamic intrusion detection methods may lose this information in separated time gaps, due to the employed time-window or snapshots-based methods.
\begin{equation}
    \left\{\begin{array}{l}
\partial_t\mathbf{X}=\mathbf{F}\left( \mathbf{X}, \mathbf{\Theta}_{}\right) \\
\mathbf{X}_{0}=\mathbf{C},
\end{array} \right.
\end{equation}
where $\mathbf{F}$ is a matrix–valued nonlinear function conditioned on graph $\mathcal{G}$, and $\mathbf{\Theta}$ is the tensor of trainable parameters. 


Specifically, we aim to amplify the important dimensions of disentangled representations and depress the influence of trivial feature elements in the diffusion process. To formulate this process, we consider PM (PeronaMalik) diffusion \cite{perona1990scale}, a type of nonlinear filtering. It can be expressed as:
 \begin{equation}
  \left\{\begin{array}{ll}
    \frac{\partial x(u,t)}{\partial t} & = \operatorname{div}[g(|\nabla x(u, t)|) \nabla x(u, t)] \\  
    x(u, 0)& = c,
    \end{array}\right.
\end{equation}
where $div$ is the divergence operator, $\nabla$ is the gradient operator, and $g$ is a function inversely proportional to the absolute value of the gradient. 

In addition, the edges in dynamic multi-layer graphs always have different timestamps and are in different layers. There are also different interactions at different layers and times. These factors will dynamically affect the information exchange results over graphs. Therefore, we aim to incorporate the above factors into the diffusion process. We propose the following layer-temporal coefficient $s_{ij}$ between nodes i,j in $t_{ij}$ time: 

\begin{align}
    s_{ij}&=f(l_i||l_j||\phi(t-t_{ij})),\\
    f(\mathbf{x})&= \mathbf{W}^{(2)} \cdot \operatorname{ReLU}\left(\mathbf{W}^{(1)} \mathbf{x}\right),
\end{align}
where $f(\cdot)$ is defined in (15), $\phi(\cdot)$ represents a generic time encoder \cite{xu2020inductive}, $\mathbf{W}^{(1)}$  and  $\mathbf{W}^{(2)}$  are the parameters of the first layer and second layer MLP.

To transfer the above continuous PM diffusion to the multi-layer graph, we need to define the differential operators on the multi-layer graph. As known from previous literature \cite{chung1997spectral}, the gradient operator corresponds to the instance matrix $\mathbf{M}$, while the divergence operator corresponds to the matrix $\mathbf{M}^{\top}$, and we can compute the matrix $M$ by the equation $\mathbf{M}^T\mathbf{M}=\mathbf{D}-\mathbf{A}$, where $\mathbf{D}$ is the diagonal matrix.
Then our novel multi-layer diffusion can be expressed as:

\begin{equation}
     {\partial\mathbf{X}_t}=-\mathbf{M}^{\top} \sigma(\mathbf{M} \mathbf{X}\mathbf{K}^{\top})\mathbf{S}\left(\mathbf{M} \mathbf{X} \mathbf{K}^{\top}\right) \mathbf{K},
     \label{eqf}
\end{equation}
where $\mathbf{K}$ is a transformation matrix, $\mathbf{S}$ is the layer-temporal influence coefficients and $\sigma(x)$ represents the function $\exp(-|x|)$. 
The solution to equation \ref{eqf} can be expressed as:
\begin{equation}
  \mathbf{X}_{t+\Delta t}=\mathbf{X}_t+\int_t^{t+\Delta t}{\partial_t \mathbf{X}_t}d \tau,
  \label{eqk}
\end{equation}
where $t$ is the last edge occurrence time and $\Delta t $ is the edge
duration and we use the Runge-Kutta method to solve this equation.

\subsection{Classifier and Loss Function}
Finally, we make the two-step predictions for intrusion detection. We utilize the first MLP to classify whether the traffic is benign or anomalous and utilize the second MLP to detect the specific type of attack. When an unknown attack invades, the direct multi-classifications will be easy to fail to assign the anomalous label thus leading to poor performance. In contrast, through the first-step binary classification, 3D-IDS will focus more on inconsistencies with normal behavior to improve the performance of detecting unknown attacks. The following second multi-classification will further alert the administrators that what kind of attack it is more similar to so that similar mitigation measures can be taken. Specifically, the intrusion loss can be expressed as:

\begin{equation}
    \mathcal{L}_{\text {Int}}=-\sum_{i=1}^{m}(\log(1-p_{\mathrm{nor},i})+\log(p_{\mathrm{att}, i})+\sum_{i=1}^{m}\sum_{j=1}^{K}y_{i,k}{\log(p_{i,k}}).
    \label{eqg}
\end{equation}
where $m$ is the batch size, $K$ is the number of attack classes, $p_{\mathrm{nor},i}$ is the probability of normal, $p_{\mathrm{att}, i}$ is the probability of attack.
Additionally, the adjacent time intervals may cause adjacent times embedding to be farther apart in embedded space, due to the learning process independency. To address this problem, we constrained the variation between adjacent timestamps embedding by minimizing the Euclidean Distance: 
\begin{equation}
    \mathcal{L}_{\text {Smooth}}=\sum_{t=0}^{T}\left\|\mathbf{X}_{t+\Delta {t}}-\mathbf{X}_{t}\right\|_{2}.
    \label{eqh}
\end{equation}
Finally, the overall loss of 3D-IDS can be expressed as follows:
\begin{equation}
 \mathcal{L}= \mathcal{L}_{\text {Int}}+\alpha  \mathcal{L}_{\text {Smooth }}+\beta  \mathcal{L}_{\text{Dis}},
 \label{eqj}
\end{equation}
where $\alpha$, and $\beta$ are trade-off parameters.

\begin{table*}[t]
\centering
  \setlength{\belowcaptionskip}{2pt}
  \caption{Comparisons of binary classification on five datasets. The results with $\ddagger$ are directly copied from \cite{Anomal-E}.}
\label{E-1}
\resizebox{\linewidth}{!}{
\begin{tabular}{c|cc|cc|cc|cc|cc} 
\toprule
\multirow{2}{*}{Methods} & \multicolumn{2}{c|}{CIC-TON-IoT}   & \multicolumn{2}{c|}{CIC-BoT-IoT}    & \multicolumn{2}{c|}{EdgeIIoT}  &\multicolumn{2}{c|}{NF-UNSW-NB15-v2} &\multicolumn{2}{c}{NF-CSE-CIC-IDS2018-v2}                                   \\
     & F1          & AUC          & F1           & AUC               & F1            & AUC               & F1              & AUC               & F1                 & AUC     \\  
\midrule 
TGN \cite{rossi2020temporal}   & \underline{$89.90$$\pm$\scriptsize$1.66$} & \underline{$82.09$$\pm$\scriptsize$1.36$}  & $96.84$$\pm$\scriptsize$0.44$  & $94.41$$\pm$\scriptsize$0.81$ & \underline{$94.99$$\pm$\scriptsize$0.61$} & $89.50$$\pm$\scriptsize$2.04$  & $93.55$$\pm$\scriptsize$0.23$ & $88.01$$\pm$\scriptsize$1.97$  & $95.11$$\pm$\scriptsize$0.46$         & \underline{$91.30$$\pm$\scriptsize$0.76$}      \\
EULER \cite{kingeuler}         &$89.73$$\pm$\scriptsize$1.13$    & $80.48$$\pm$\scriptsize$2.46$  & $96.00$$\pm$\scriptsize$0.29$   &  $91.47$$\pm$\scriptsize$1.36$  & $92.89$$\pm$\scriptsize$0.32$  & \underline{$90.64$$\pm$\scriptsize$1.80$}   & $92.76$$\pm$\scriptsize$0.86$     & $86.97$$\pm$\scriptsize$1.11$         & $95.87$$\pm$\scriptsize$0.51$         & $90.76$$\pm$\scriptsize$0.64$    \\
AnomRank \cite{2019Fast}             & $76.51$$\pm$\scriptsize$0.98$   & $77.41$$\pm$\scriptsize$1.64$  & $84.84$$\pm$\scriptsize$0.49$   & $82.50$$\pm$\scriptsize$0.59$        & $78.37$$\pm$\scriptsize$0.43$    & $81.36$$\pm$\scriptsize$0.41$        & $90.54$$\pm$\scriptsize$2.40$     & $79.63$$\pm$\scriptsize$0.12$        & $90.08$$\pm$\scriptsize$0.82$          & $83.76$$\pm$\scriptsize$0.35$\\
DynAnom \cite{guo2022dynanom}        & $79.23$$\pm$\scriptsize$1.81$   & $75.22$$\pm$\scriptsize$0.92$  & $83.25$$\pm$\scriptsize$0.62$   & $79.04$$\pm$\scriptsize$0.84$        & $81.56$$\pm$\scriptsize$0.94$    & $83.94$$\pm$\scriptsize$0.36$        & $89.11$$\pm$\scriptsize$1.48$     & $85.25$$\pm$\scriptsize$0.64$        & $91.21$$\pm$\scriptsize$0.95$          & $88.79$$\pm$\scriptsize$0.54$ \\
\midrule
Anomal-E \cite{Anomal-E}&-  & -  &  - &     -  & -   &    -    & $91.89\ddagger$     &-       & $94.51\ddagger$        &- \\
GAT \cite{velivckovic2017graph}     & $86.30$$\pm$\scriptsize$1.16$   & $74.66$$\pm$\scriptsize$1.37$  & $94.56$$\pm$\scriptsize$0.75$   & $93.09$$\pm$\scriptsize$2.83$        & $93.30$$\pm$\scriptsize$0.14$    &$88.30$$\pm$\scriptsize$1.56$         & $92.20$$\pm$\scriptsize$1.60$     & $89.91$$\pm$\scriptsize$0.62$  & \underline{$96.08$$\pm$\scriptsize$0.24$}    & $90.56$$\pm$\scriptsize$0.34$    \\
E-GraphSAGE \cite{lo2022graphsage}  & $89.46$$\pm$\scriptsize$1.25$   & $79.56$$\pm$\scriptsize$1.63$  & $93.74$$\pm$\scriptsize$0.76$   & $90.53$$\pm$\scriptsize$1.90$       & $92.10$$\pm$\scriptsize$1.46$    & $89.10$$\pm$\scriptsize$0.64$ &\underline{$94.10$$\pm$\scriptsize$0.33$} &\underline{$90.39$$\pm$\scriptsize$0.26$} & $95.71$$\pm$\scriptsize$0.35$        & $90.22$$\pm$\scriptsize$0.48$ \\
DMGI \cite{park2020unsupervised}    & $88.83$$\pm$\scriptsize$0.48$   & $79.13$$\pm$\scriptsize$2.11$  & $96.07$$\pm$\scriptsize$1.89$   &  $92.65$$\pm$\scriptsize$1.57$       & $93.83$$\pm$\scriptsize$1.67$    & $86.03$$\pm$\scriptsize$2.45$        & $93.11$$\pm$\scriptsize$0.98$     & $88.51$$\pm$\scriptsize$1.00$         & $93.87$$\pm$\scriptsize$0.84$         & $87.56$$\pm$\scriptsize$0.55$  \\
SSDCM \cite{mitra2021semi}  & $89.23$$\pm$\scriptsize$0.87$ & $80.84$$\pm$\scriptsize$2.32$ & \underline{$97.11$$\pm$\scriptsize$0.63$} &\underline{$94.82$$\pm$\scriptsize$0.96$} & $94.72$$\pm$\scriptsize$1.59$ & $86.69$$\pm$\scriptsize$0.76$        & $93.30$$\pm$\scriptsize$0.25$     & $89.22$$\pm$\scriptsize$1.94$         & $94.96$$\pm$\scriptsize$0.52$         & $88.61$$\pm$\scriptsize$0.38$  \\ 
\midrule
MLP \cite{roopak2019deep}      & $80.74$$\pm$\scriptsize$0.43$   & $61.80$$\pm$\scriptsize$1.48$  & $93.01$$\pm$\scriptsize$0.60$   & $87.90$$\pm$\scriptsize$0.54$        & $88.78$$\pm$\scriptsize$0.44$    & $86.00$$\pm$\scriptsize$1.49$        & $93.12$$\pm$\scriptsize$0.64$     & $89.92$$\pm$\scriptsize$0.55$         & $94.59$$\pm$\scriptsize$0.94$         & $90.42$$\pm$\scriptsize$0.89$      \\

MStream \cite{Bhatia_2021} & $73.90$$\pm$\scriptsize$1.13$   & $70.22$$\pm$\scriptsize$1.61$  & $78.48$$\pm$\scriptsize$0.19$   & $74.04$$\pm$\scriptsize$1.66$     & $82.47$$\pm$\scriptsize$1.67$    & $77.89$$\pm$\scriptsize$0.58$        & $89.47$$\pm$\scriptsize$1.13$   & $84.38$$\pm$\scriptsize$1.01$       & $88.34$$\pm$\scriptsize$0.45$ & $83.66$$\pm$\scriptsize$1.79$\\

LUCID \cite{doriguzzi2020lucid} & $83.62$$\pm$\scriptsize$1.69$   & $72.31$$\pm$\scriptsize$1.14$  & $94.36$$\pm$\scriptsize$0.41$   & $89.46$$\pm$\scriptsize$0.72$     & $88.94$$\pm$\scriptsize$1.73$    & $85.23$$\pm$\scriptsize$0.94$        & $92.77$$\pm$\scriptsize$1.39$   & $88.32$$\pm$\scriptsize$0.91$       & $95.84$$\pm$\scriptsize$1.46$ & $90.75$$\pm$\scriptsize$0.79$\\
\midrule
\textbf{Ours (3D-IDS)} &{$\textbf{91.57}$$\pm$\scriptsize$\textbf{0.40}$}&{$\textbf{84.06}$$\pm$\scriptsize$\textbf{1.01}$}&{$\textbf{98.24}$$\pm$\scriptsize$\textbf{0.32}$}&{$\textbf{96.32}$$\pm$\scriptsize$\textbf{0.25}$}&{$\textbf{96.83}$$\pm$\scriptsize$\textbf{0.36}$}&{$\textbf{92.34}$$\pm$\scriptsize$\textbf{1.10}$}&{$\textbf{95.45}$$\pm$\scriptsize$\textbf{0.67}$}&{$\textbf{91.55}$$\pm$\scriptsize$\textbf{1.03}$}&{$\textbf{96.34}$$\pm$\scriptsize$\textbf{0.21}$}&{$\textbf{93.23}$$\pm$\scriptsize$\textbf{1.50}$}  \\
\bottomrule
\end{tabular}}
\end{table*}
\section{Experiments}


\subsection{Experimental Settings}
\textbf{Datasets\iffalse\footnote{More  details are available in Table \ref{DATASETTABLE} of Appendix \ref{sec-datasets}.}\fi:} We conduct experiments on five popular datasets that involve massive network traffic over the internet of things (IoT). We give detailed descriptions as follows:
\begin{itemize}
  \item CIC-ToN-IoT \cite{moustafa2019new}: This dataset is generated from the existing ToN-IoT dataset \cite{moustafa2021new} by a network traffic tool CICFlowMeter \cite{lashkari2017cicflowmeter}, where TON-IoT is a well-known database for intrusion detection collected from Telemetry datasets \cite{alsaedi2020ton_iot} of IoT services.  This dataset consists of $5,351,760$ flows, with $53.00\%$ attack samples and $47.00\%$ benign samples.

  \item CIC-BoT-IoT \cite{sarhan2021explainable}: It is generated from the existing BoT-IoT dataset \cite{koroniotis2019towards} by CICFlowMeter. This dataset consists of $6,714,300$ traffic flows, with $98.82\%$ attack samples and $1.18\%$ benign samples.
  \item EdgeIIoT \cite{ferrag2022edge}: It is collected from an IoT/IIoT system that contains mobile devices and sensors. This dataset includes $1,692,555$ flows, with $21.15\%$ attack samples and $78.85\%$ benign samples.
\item NF-UNSW-NB15-v2 \cite{sarhan2022towards}: It is NetFlow-based and generated from the UNSW-NB15 \cite{moustafa2015unsw} dataset, which has been expanded with additional NetFlow features and labeled with respective attack categories. This dataset includes $2,390,275$ flows, with $3.98\%$ attack samples and $96.02\%$ benign samples.
\item NF-CSE-CIC-IDS2018-v2 \cite{sarhan2022towards}: It is a  NetFlow-based dataset generated from the original pcap files of CSE-CIC-IDS2018 \cite{sharafaldin2018toward} dataset. This dataset includes $18,893,708$ flows, with $11.95\%$ attack samples and $88.05\%$ benign samples.
\end{itemize}
\textbf{Configurations\iffalse\footnote{More details are available in Appendix \ref{sec:Implementation Details}.}\fi:} 
All experiments and timings are conducted on a machine with Intel Xeon Gold 6330@ $2.00$GHz, RTX$3090$ GPU, and $24$G memory. We use the Adam optimizer with a learning rate of $0.01$, the learning rate scheduler reducing rate as $0.9$, with weight decay being $1e^{-5}$. We train all the models with 500 epochs. \\
\textbf{Baselines\iffalse\footnote{More  details are available in Appendix \ref{A:Baselines}.}\fi:} 
To evaluate the performance of the proposed 3D-IDS, we select $10$ deep learning based-models as baselines, including $3$ sequence models (i.e., MLP \cite{roopak2019deep}, MStream \cite{Bhatia_2021}, LUCID \cite{doriguzzi2020lucid}), $4$ static GCN models (i.e., GAT \cite{velivckovic2017graph} and E-GraphSAGE \cite{hamilton2017representation}, SSDCM \cite{mitra2021semi}, DMGI \cite{park2020unsupervised}), where SSDCM and DMGI are designed for static multi-layer graphs,  $4$ dynamic GCN models (i.e., TGN \cite{rossi2020temporal}, EULER \cite{kingeuler}, AnomRank \cite{2019Fast}, DynAnom \cite{2019Fast}). Additionally, we choose $3$ rule-based baselines to compare with the proposed 3D-IDS, (i.e., ML \cite{sarhan2021netflow}, AdaBoost \cite{lalouani2021robust}, and Logistic Regression).

\noindent
\textbf{Metrics:} We follow the previous works \cite{sarhan2022evaluating} to evaluate the performances of all baselines by two commonly used metrics in intrusion detection including F1-score (F1) and ROC-AUC score (AUC).\\
\vspace{-20pt}



\begin{figure*}[htb]
    \centering
    \subfigbottomskip=-3pt 
    \subfigcapskip=-7pt 
    \setlength{\abovecaptionskip}{8pt}  
    \includegraphics[width=0.90\linewidth,height=0.4cm]{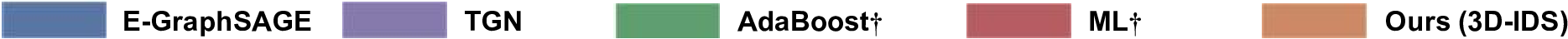}
    \subfigure[EdgeIIoT]{
    \includegraphics[width=0.53\linewidth]{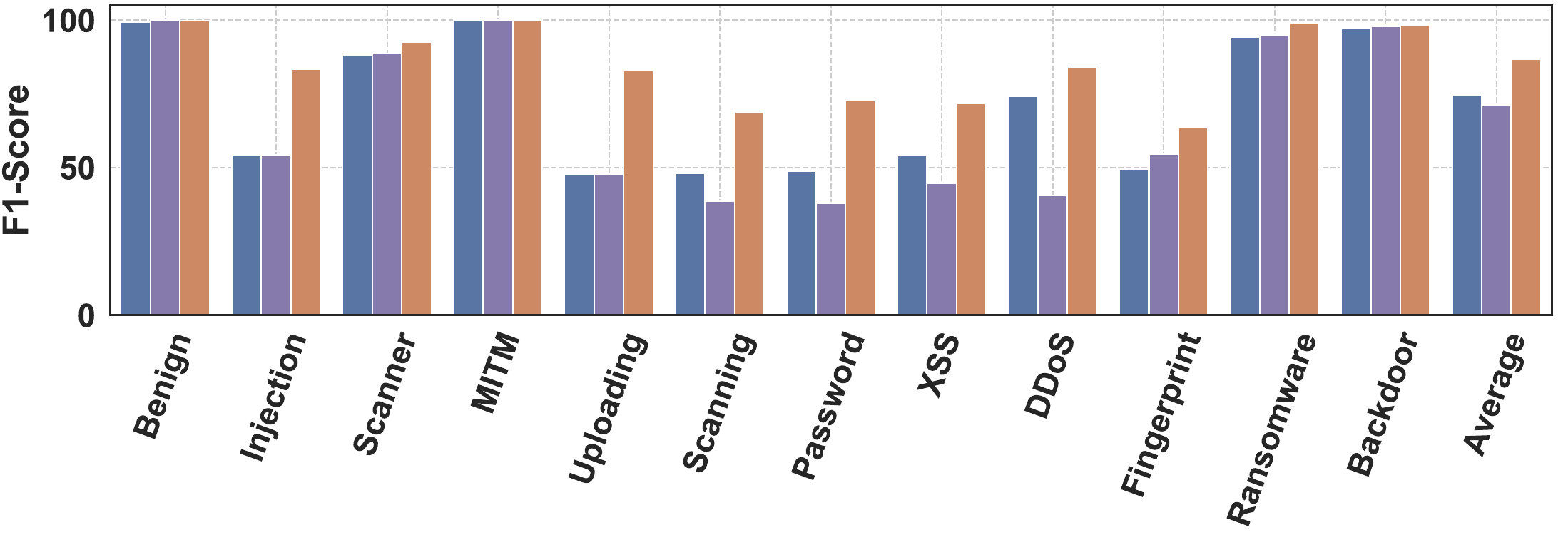}
    }
    \subfigure[CIC-ToN-IoT]{
    \includegraphics[width=0.44\linewidth]{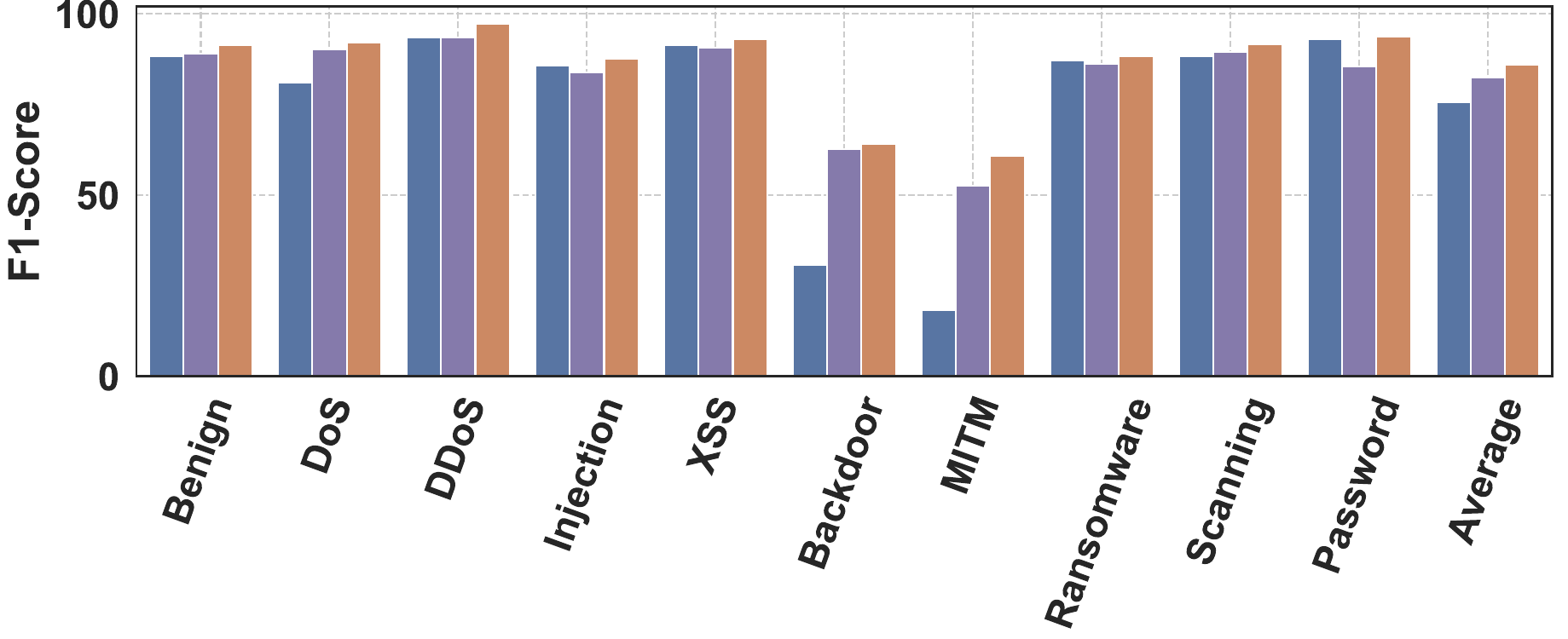}
    }
    \subfigure[CIC-BoT-IoT]{
    \includegraphics[width=0.24\linewidth]{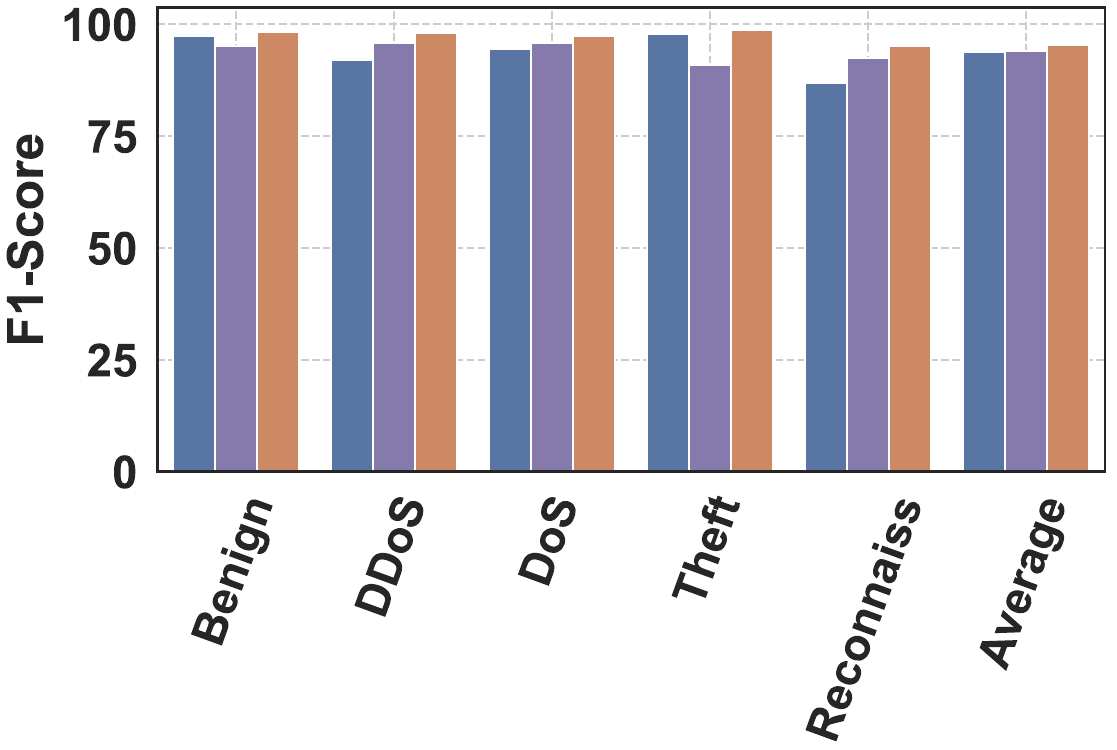}
    }
    \subfigure[NF-UNSW-NB15-v2]{
    \includegraphics[width=0.315\linewidth]{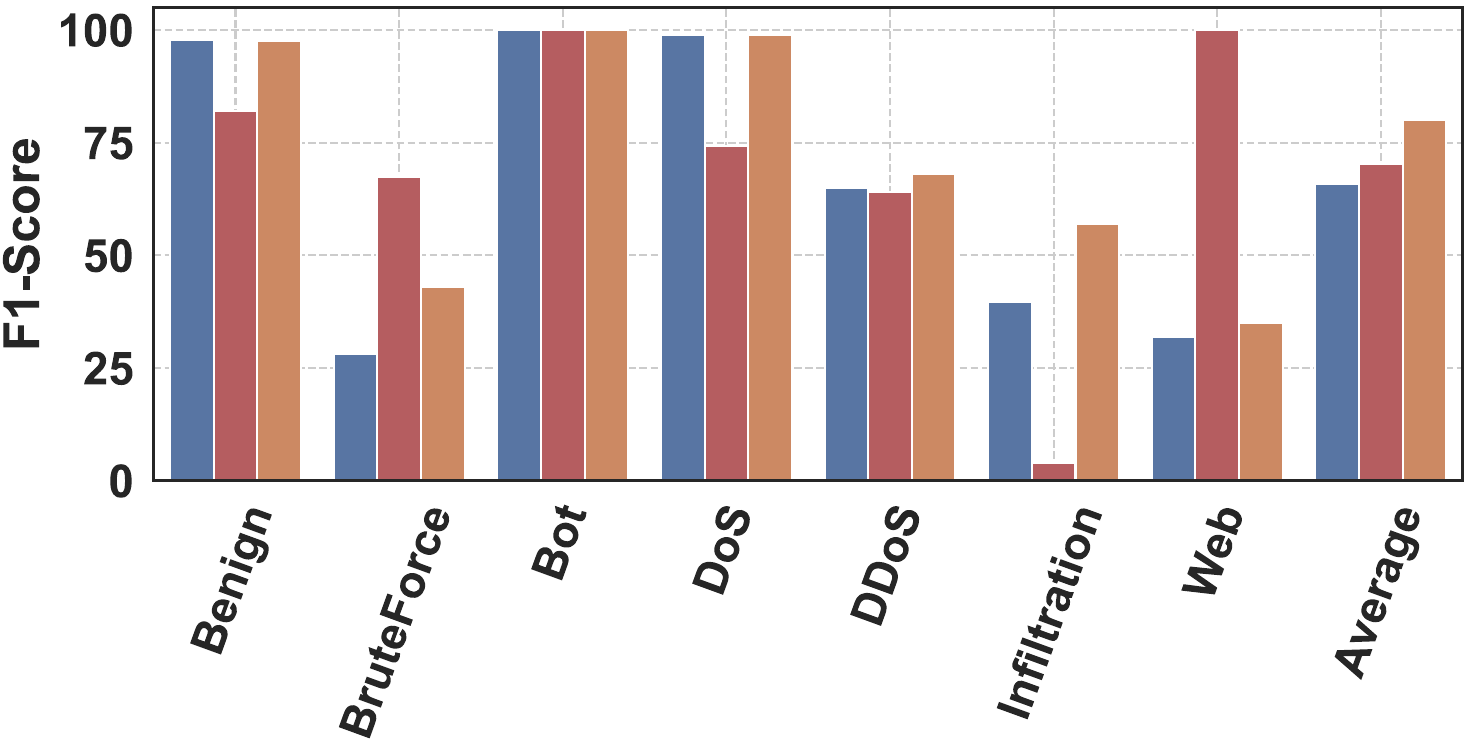}
    }
    \subfigure[NF-CSE-CIC-IDS2018-v2]{
    \includegraphics[width=0.40\linewidth]{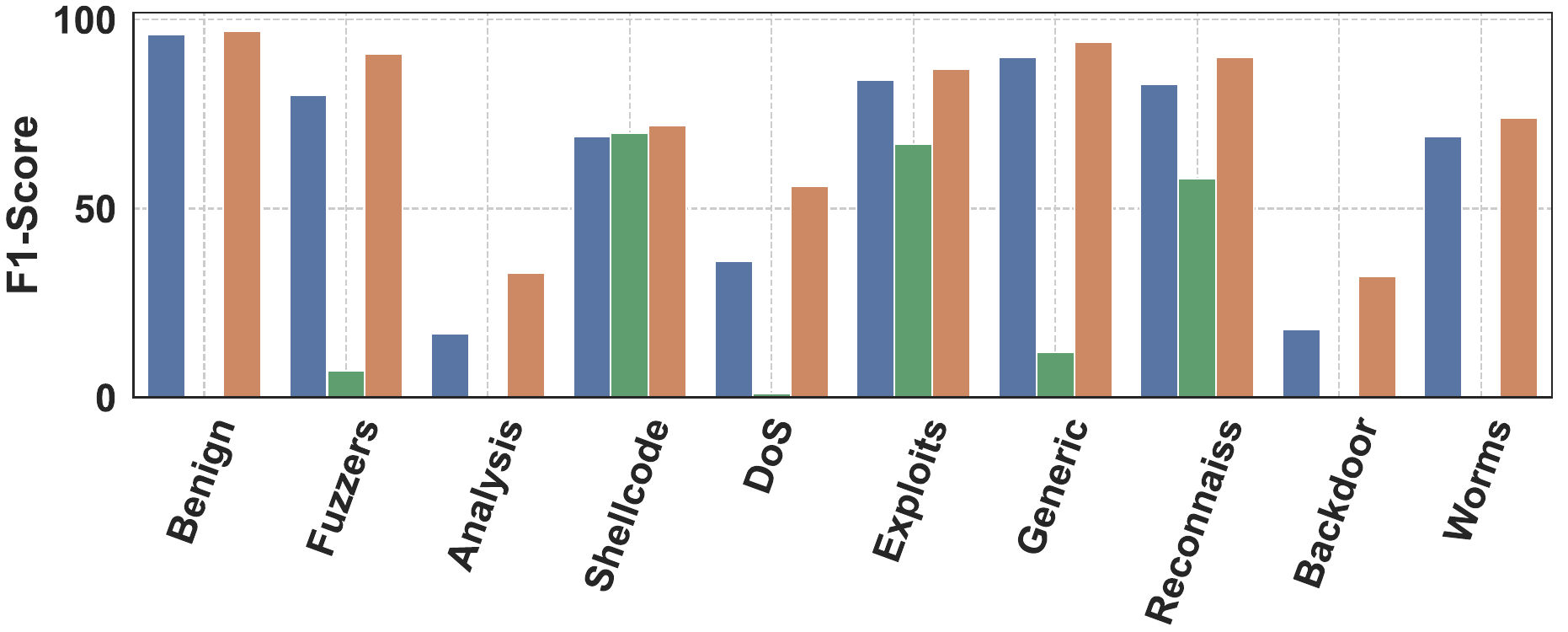}
    }
    \caption{Comparisons of multi-classification. Here $\dagger$ indicates that the results are directly copied from the previous works.}
    \label{classfication}
\end{figure*}

\hspace{-10pt}
\subsection{Main Results}

\subsubsection{Comparisons of binary classification.}
Under this setting, we classify a traffic flow as an attack or a benign one. We categorize the baselines into three groups, including dynamic GCNs at the top of Table \ref{E-1}, static GCNs at the middle of the table, and another three baselines at the bottom. It should be noted that AnomRank and DynAnom are two popular baselines for anomaly detection. We run our experiment $5$ times and report the mean and variance values. The comparison results in Table \ref{E-1} show that our 3D-IDS consistently performs the best among all baselines over the five benchmarks, which shows the superiority of our method for intrusion detection. Specifically, compared to the E-GraphSAGE, the previous state-of-art GCN-based approach, our method achieves a $4.80\%$ higher F1-score over the CIC-BoT-IoT dataset. Our 3D-IDS outperforms AnomRank, the previous state-of-the-art method for anomaly detection on F1, by $15.27$ points over the EdgeIIoT dataset. We attribute the above results to the gains of our statistical disentanglement, representational disentanglement and dynamic graph diffusion method. 

\subsubsection{Comparisons of multi-classification.}
We compare the performance of our method to four baselines in declaring the specific attack type. These baselines include E-GraphSAGE \cite{lo2022graphsage}, TGN \cite{rossi2020temporal}, ML \cite{wang2016machine}, and AdaBoost \cite{schapire2013explaining}, as they are representative of different types of intrusion detection models and have been widely used in previous studies. Figure \ref{classfication} shows that the proposed 3D-IDS consistently achieves the best results among all others over five datasets. For example, ours yields higher classification accuracy of up to $25$ points compared to the baseline TGN for detecting  Injection attacks. The existing graph-based methods, including E-GraphSAGE, ML, and AdaBoost, perform inconsistently in the identification of complex attacks (e.g., MITM, Uploading, and XSS). We also observe that some attacks that are not easily detected by the baseline approaches, can be identified by the proposed 3D-IDS with high F1 scores. For example, E-GraphSAGE only achieves $18.34\%$ and $30.7\%$ F1 scores on CIC-ToN-IoT for MITM and Backdoor attacks, respectively, while our 3D-IDS is able to obtain higher than $23\%$ F1 score for each attack. These results further show the superiority of the proposed 3D-IDS. We also find that the average scores of our method on CIC-BoT-loT and CIC-ToN-loT are lower than the ones on the other three datasets. The underlying reason is the unbalanced attack distributions in the training set, where the dominant type may mislead the classifications. Such a finding aligns with previous work in the field of computer vision \cite{fisher2017operational}. Nevertheless, the proposed 3D-IDS is still the best under such distributions. We leave this interesting observation as our future work.

\subsubsection{Comparisons of unknown attacks.}
To further investigate the performance of detecting unknown attacks, we conduct experiments on the four attack types by discarding the corresponding instances in the train set and detecting them in the test set. We run our experiments $5$ times and report the mean and variance values with different random seeds. Table \ref{UAC} reports the classification results on the CIC-ToN-IOT dataset. It shows that the statistical rule-based method Logistic Regression can only achieve as low as a $1.68$ F1 score for DDoS attacks, this confirms our analysis at the very beginning that rule-based methods can hardly detect unknown attacks. The score of graph-based E-GraphSAGE is much smaller than 3D-IDS, e.g., $6.05\%$ for MITM, indicating the limitations of the static graph in detecting unknown attacks. We also observe that TGN performs better than E-GraphSAGE, although both of them are graph-based methods. We attribute the improvement to the dynamic module for TCN. Nevertheless, our 3D-IDS outperforms all these methods by a large margin, with an average score of $33.65\%$ on the four attacks. The results also suggest that our method is more consistent in detecting various unknown attacks, showing the effectiveness of the two disentanglements.

\begin{table}

\centering
\caption{Unknown classification on the CIC-ToN-IoT dataset with metric F1-score (\%)}
\label{UAC}
\resizebox{\linewidth}{!}{

\begin{tabular}{cccccc} 
\toprule
Method                             &Logistic Regression         & MSteam      & E-GraphSAGE                   & TGN                     & \textbf{Ours (3D-IDS)}     \\ 
\midrule \midrule
DDoS                          & $1.68$$\pm$\scriptsize$0.67$  & $26.73$$\pm$\scriptsize$1.21$  & $10.20$$\pm$\scriptsize$1.46$                    & $32.84$$\pm$\scriptsize$1.03$              & {$\textbf{41.78}$$\pm$\scriptsize$\textbf{0.26}$}  \\
MITM                          & $2.17$$\pm$\scriptsize$0.84$  & $12.82$$\pm$\scriptsize$1.90$  & $6.05$$\pm$\scriptsize$0.35$                     & $15.32$$\pm$\scriptsize$0.54$              & {$\textbf{34.91}$$\pm$\scriptsize$\textbf{0.91}$}  \\
Injection                     & $0.00$$\pm$\scriptsize$0.34$  & $15.73$$\pm$\scriptsize$1.73$  & $12.37$$\pm$\scriptsize$0.88$                    & $22.83$$\pm$\scriptsize$0.35$              & {$\textbf{25.63}$$\pm$\scriptsize$\textbf{0.93}$}  \\
Backdoor                      & $3.13$$\pm$\scriptsize$1.33$  & $20.85$$\pm$\scriptsize$0.63$  & $9.51$$\pm$\scriptsize$0.46$                     & $23.10$$\pm$\scriptsize$1.03$              & {$\textbf{32.29}$$\pm$\scriptsize$\textbf{0.81}$}  \\
\bottomrule
\end{tabular}}
\vspace{-5pt}
\end{table}

\subsection{Ablation Study}

In this section, we conduct an ablation study on the CIC-ToN-IoT dataset to evaluate the effectiveness of each component. We remove our statistical disentanglement and denote it as "w/o SD". We use "w/o RD" and "w/o MLGRAND"  to refer to the model that removes representational disentanglement and the multi-Layer graph diffusion module, respectively. Table \ref{Tab:abl} reports the comparison results. It shows that removing the multi-Layer graph diffusion module leads to the most significant performance degradation, e.g., an $18.33$ points decrease in AUC, indicating that it is the key component for the accuracy of the proposed 3D-IDS. 
Our second disentangled memory is also non-trivial to the overall detection accuracy, as removing this component can decrease the performance by $12.47$ points in AUC. We observe that the multi-layer graph diffusion module also benefits the model performance. The above ablation study further confirms the effectiveness of the three key components.

\begin{table}[!htb]
  \centering
  \caption{Ablation study of 3D-IDS. F1-score (\%), Precision (\%), ROC-AUC (\%), Recall (\%), all metrics above are the average of five repeated experiments on the CIC-ToN-IoT dataset.}
  \resizebox{\linewidth}{!}{
  \begin{tabular}{c|cccc} 
  \toprule
  Variants              & P & R & F1 & AUC   \\ 
  \midrule \midrule
  w/o SD  & $92.70$$\pm$\scriptsize$0.46$  & $90.71$$\pm$\scriptsize$0.89$ & $91.69$$\pm$\scriptsize$0.33$   & $86.87$$\pm$\scriptsize$0.54$   \\
  w/o RD   & $91.06$$\pm$\scriptsize$0.57$  & $87.32$$\pm$\scriptsize$0.67$ & $89.15$$\pm$\scriptsize$0.42$   & $83.57$$\pm$\scriptsize$1.33$   \\
  w/o MLGRAND        & $88.76$$\pm$\scriptsize$0.54$  & $84.43$$\pm$\scriptsize$0.26$ & $86.54$$\pm$\scriptsize$0.71$   & $79.32$$\pm$\scriptsize$0.30$   \\
  \hline
  \textbf{3D-IDS(ours)} &$\textbf{97.78}$$\pm$\scriptsize$\textbf{0.32}$  & $\textbf{98.06}$$\pm$\scriptsize$\textbf{0.43}$ & $\textbf{97.92}$$\pm$\scriptsize$\textbf{0.26}$   & $\textbf{96.04}$$\pm$\scriptsize$\textbf{0.25}$   \\
  \bottomrule
  \end{tabular}}
   \label{Tab:abl}
  \end{table}
  \vspace{-10pt}
\subsection{Discussion}
\begin{figure}
	\centering  
        \setlength{\abovecaptionskip}{8pt}  
	\subfigure[Origin]{
		\includegraphics[width=0.48\linewidth]{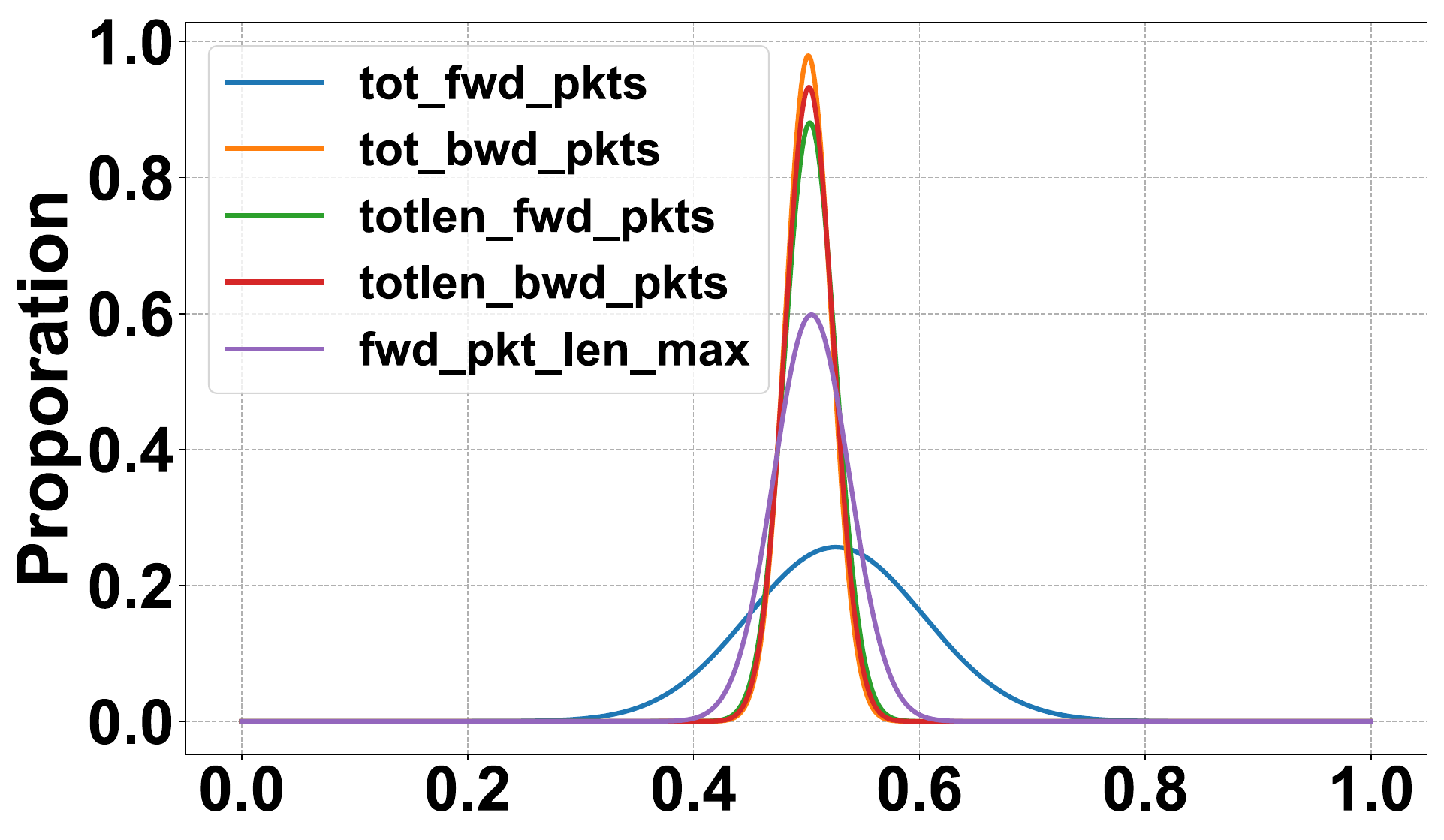}}
	\subfigure[First Distangled]{
		\includegraphics[width=0.48\linewidth]{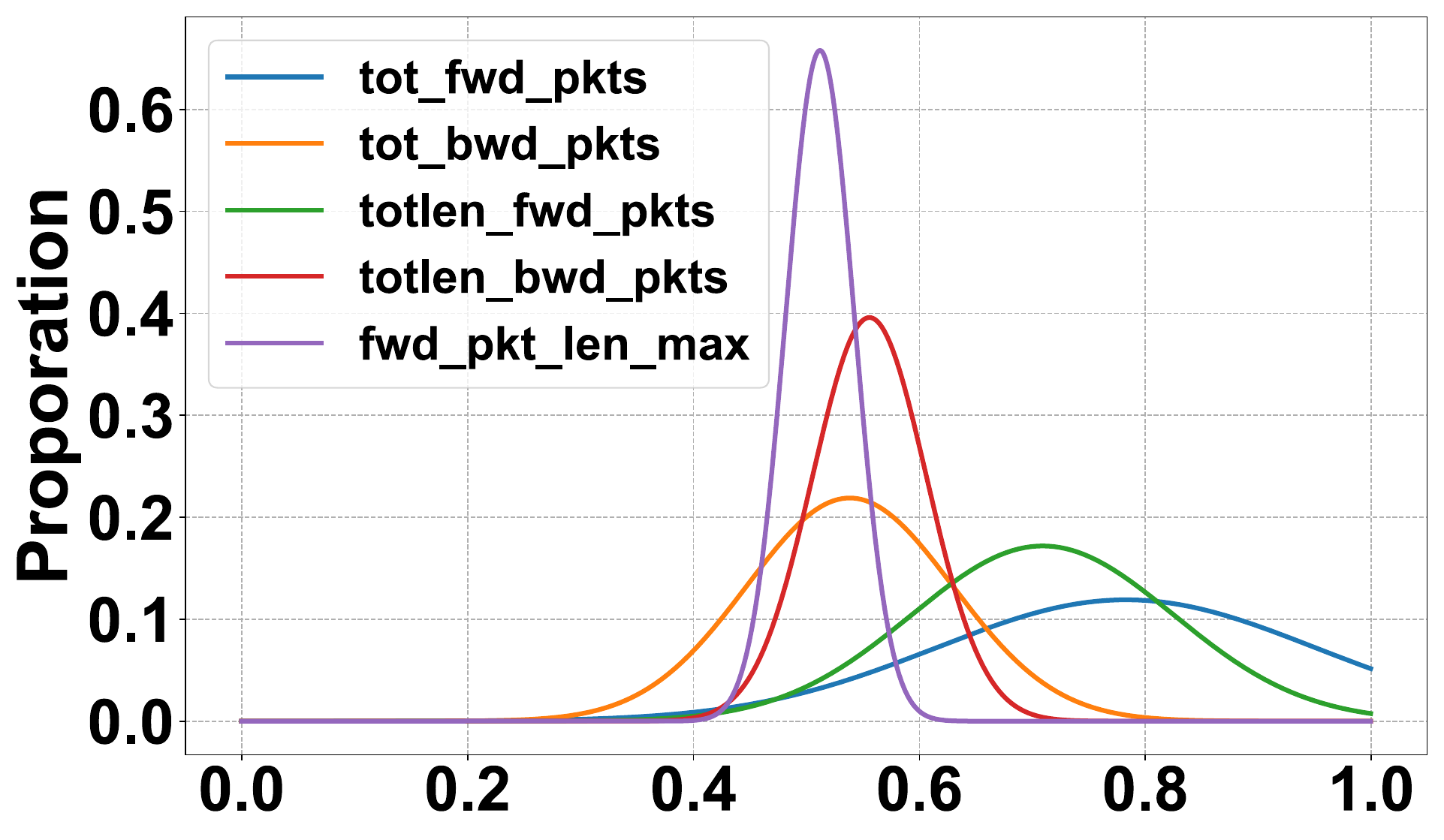}}
          \\
   \caption{Statistical disentanglement of traffic features.}
   \vspace{-5pt}
\label{dis}
\end{figure}

\textbf{RQ$1$: How does the statistical disentanglement help the detection of various attacks?}
To answer this question, we visualize the distributions of features before and after the statistical disentanglement.
Figure \ref{dis} shows the visualizations of the two distributions respectively.
We can observe that there is less overlap between distributions of features after the disentanglement compared with the original data, which demonstrates this module could decrease the mutual reference between features and enable them to be distinguishable.
We also observe that the distributions gradually shift to the right side, representing the order-preserved constraints within our disentangling method.

\begin{figure}
     \centering 
      \subfigcapskip=-3pt 
     \setlength{\abovecaptionskip}{2pt}  
\subfigure[3D-IDS]{
 \includegraphics[width=0.47\linewidth]{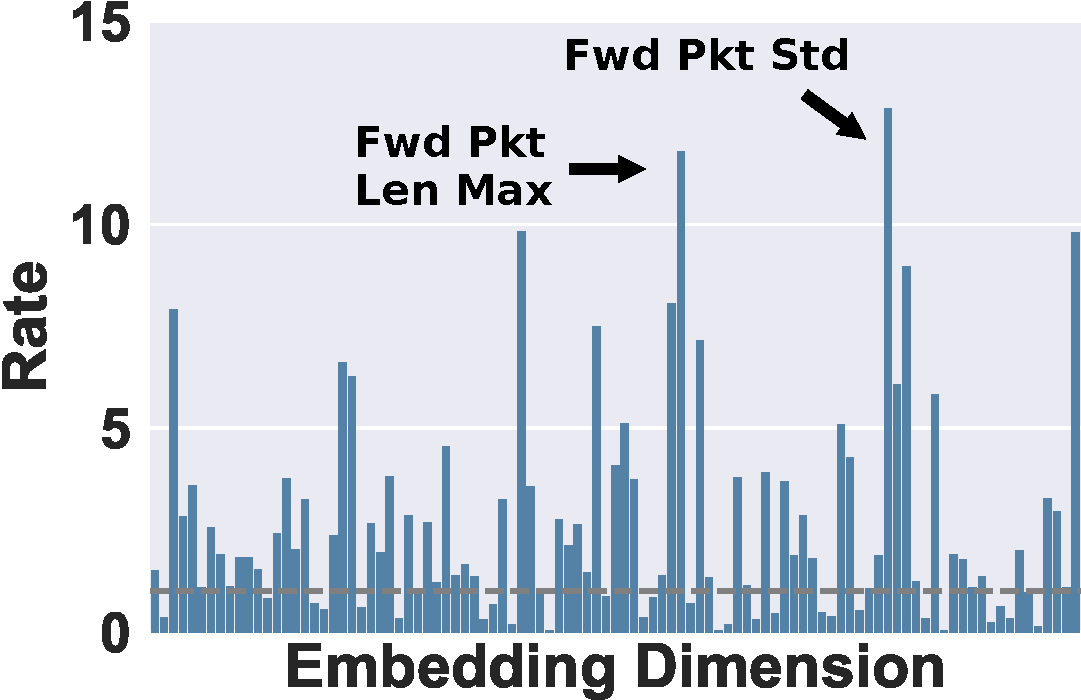}
 }
 \subfigure[3D-IDS]{
 \includegraphics[width=0.47\linewidth]{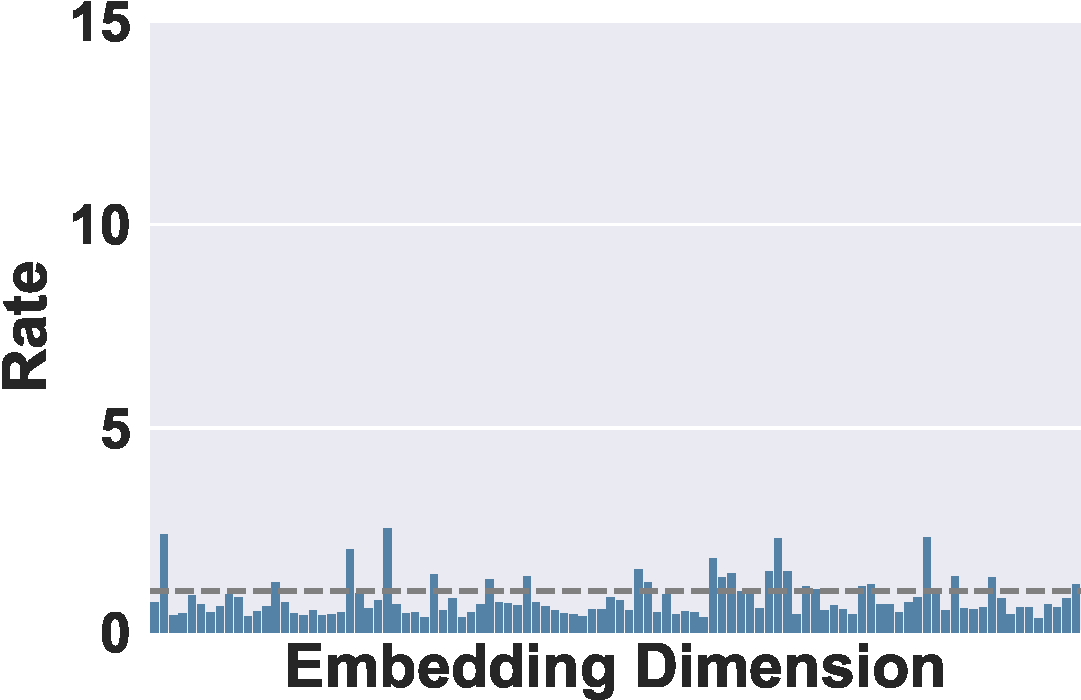}
 }
\caption{The comparison of node representation  of the Injection attack after graph aggregation of our 3D-IDS and E-GraphSAGE. The grey line presents benign data.}
\label{casestudy}
\end{figure}

\noindent
\textbf{RQ$2$: How does the representational disentanglement benefit traffic classification for a specific attack?}
To answer this question, we track several Injection attack data in the CIC-ToN-IoT dataset and obtain the representation of these data in 3D-IDS and E-GraphSAGE. Meanwhile, we calculate the average of the embedding of the benign data in the two models as the scaling ratio. As shown in Figure \ref{casestudy}, the representation values of E-GraphSAGE are much closer to the normal, which increases the probability of identifying the Injection attack as normal. It illustrates that as nodes aggregate, the discrepancies in features become blurred. The representation values of anomalous nodes gradually converge to normal, leading to incorrect link predictions. Values of our model in each dimension still deviate significantly from the normal values since ours imposes restrictions on the aggregation of information. The result proves the effectiveness of our model in maintaining a disentangled structure during the aggregation process, ensuring the presence of discrepancies, and leading to more accurate traffic classification for a specific attack. 
\begin{figure}
	\centering  
        \setlength{\abovecaptionskip}{5pt}  
	\includegraphics[width=0.95\linewidth]{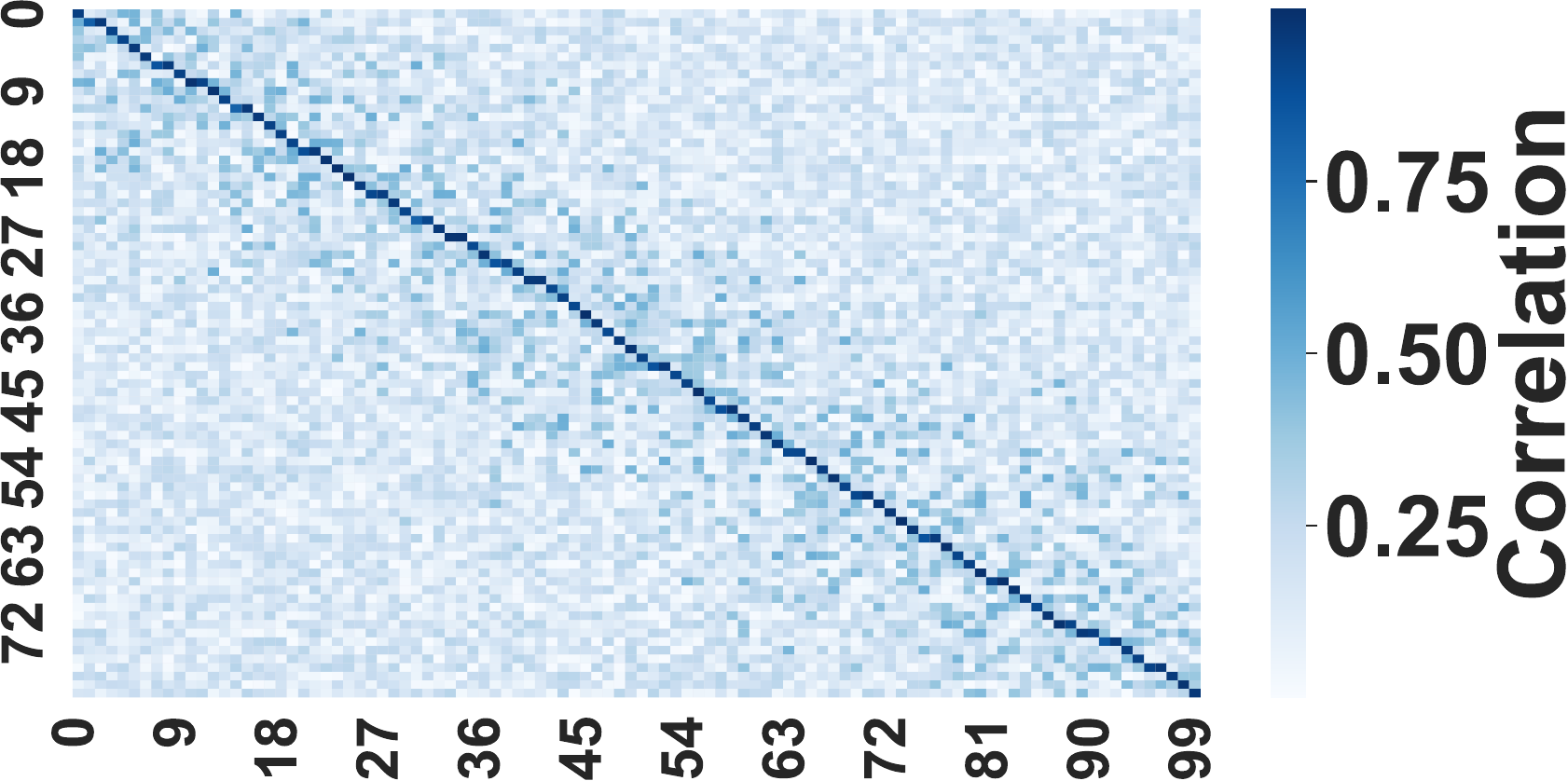}
   \caption{Linear relationship between the disentangled edge vector and the disentangled node representation.}
 \label{seconddis}
\end{figure}

\noindent
\textbf{RQ$3$: How does the multi-layer diffusion module perform effectively for intrusion detection?}
We have illustrated the principle of multi-layer diffusion in Section \ref{Multi-Layer Graph Diffusion module}. In this part, we take the MITM attack as an example to illustrate the effectiveness of spatial-temporal in intrusion detection.

\begin{figure}
  \centering
  \setlength{\abovecaptionskip}{5pt}  
  \setlength{\belowcaptionskip}{-0.3cm}
  \includegraphics[width=\linewidth]{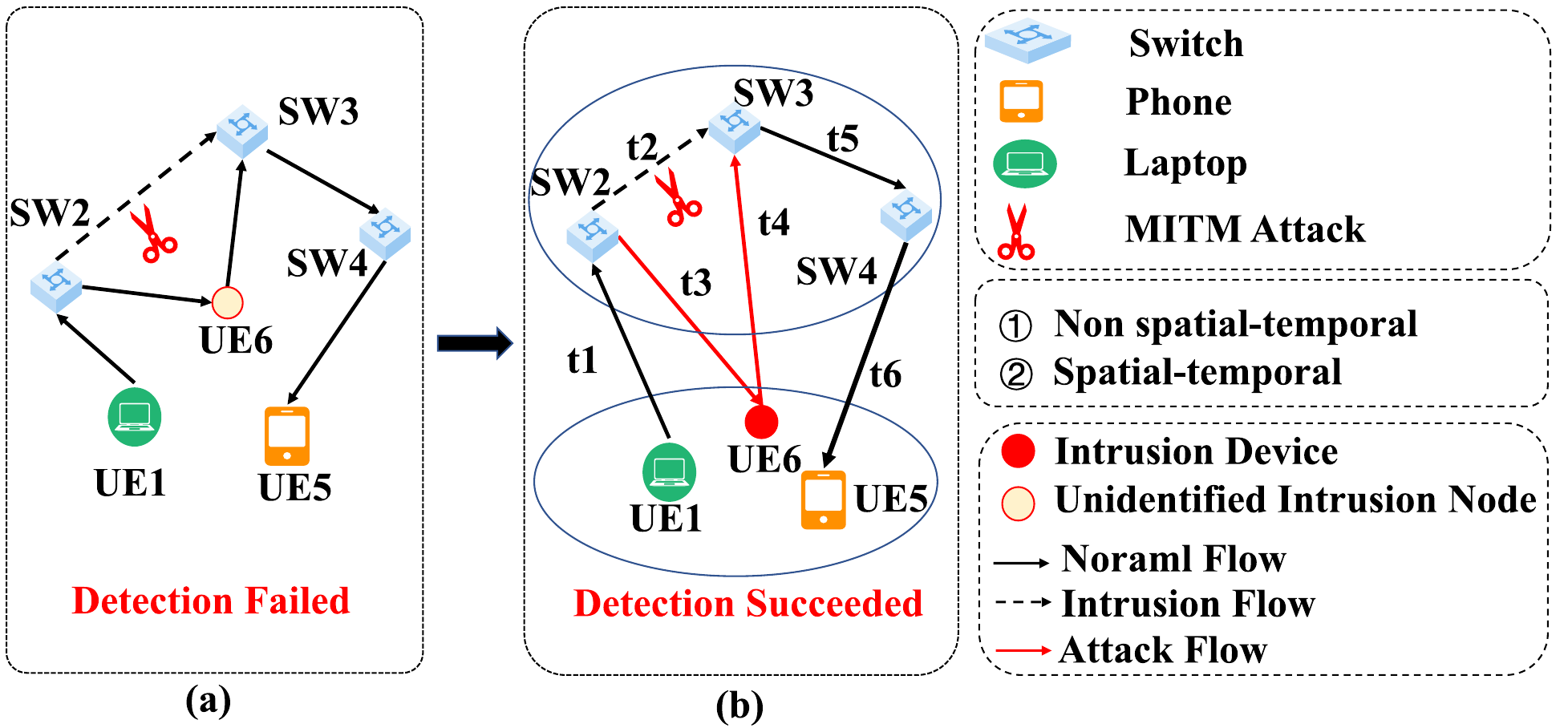}
  \caption{Spatial-temporal coupling in intrusion detection.}
  \label{attact_scenira} 
\end{figure}

Figure \ref{attact_scenira} (a) shows a deep learning-based NIDS. When a MITM attack occurs, it is difficult to detect the intrusion
since the spatial and temporal information of those packets is not considered. There are also some methods that only consider a single aspect of spatial and temporal information, such as E-GraphSAGE and MStream. In this case, for example, E-GraphSAGE mainly focuses on the spatial relationship of the set nodes and extracts features from them. However, we observe that different streams have their own timestamps from Table \ref{IPTABLE}, so the lack of temporal information makes it impossible to analyze the dynamic structural changes of the edge. Similarly,  taking the temporal information as the only effect factor will also get incomplete characteristics that do not contain spatial information (IP address). Moreover, some methods that take both the spatial and temporal information into account, such as Euler, take the snapshot method to capture the feature of the flow which does not achieve the synchronous update for spatial and  temporal information.
As shown in Figure \ref{attact_scenira} (b), intuitively, we can quickly detect that UE6 is an intrusion device of layer 1 when the flow changes from $SW2-SW3$ to $SW2-UE6$ and $UE6-SW3$ considering $SW2$, $SW3$ are layer $2$ devices. Also, we have noticed the changes in dynamic graph structure with a multi-Layer graph diffusion module to realize spatio-temporal coupling and synchronous updating. Overall, 3D-IDS performs best among these baselines in detecting various attacks.
\begin{table}
    \centering
    \caption{TimeStamp and IP}
    \label{IPTABLE}
    \resizebox{\linewidth}{!}{
    \begin{tabular}{cccc}
    \hline
        Time & TimeStamp & Src IP & Dst IP \\ \hline
        $t1$ & $25/04/2019$ $05:18:37$ pm & $183.68.192.168$ & $1.169.216.58$ \\ 
        $t2$ & $25/04/2019$ $05:18:42$ pm & $1.169.216.58$ & $25.162.192.168$ \\ 
        $t3$ & $25/04/2019$ $05:18:42$ pm & $1.169.216.58$ & $230.158.52.59$ \\ 
        $t4$ & $25/04/2019$ $05:18:49$ pm & $230.158.52.59$ & $25.162.192.168$ \\ 
        $t5$ & $25/04/2019$ $05:18:52$ pm & $25.162.192.168$ & $69.151.192.168$ \\ 
        $t6$ & $25/04/2019$ $05:19:00$ pm & $177.21.192.168$ & $230.158.52.59$ \\ \hline
    \end{tabular}}
\end{table}


\noindent
\textbf{RQ$4$: How does the two-step disentanglement facilitate the explainability of 3D-IDS?}
For this question, we rely on Figure \ref{casestudy} (a) as an example to recover the possible traffic features of a password attack. First, since the original features are retained after the double disentanglement, we can find some feature values that deviate significantly from the normal values in the node embedding. As shown in Figure \ref{seconddis}, due to the disentangled edge representation, there exists a linear relationship between the disentangled edge vector and the disentangled node representation. Therefore, based on this relationship, we observe that the deviated features "Fwd Pkt Std" and "Fwd Pkt Len Max" present a strong correlation with the password attack.

\section{CONCLUSION}
This paper quantitatively studies the inconsistent performances of existing NIDS under various attacks and reveals that the underlying cause is entangled feature distributions. These interesting observations motivate us to propose 3D-IDS, a novel method that aims to benefit known and unknown attacks with a double disentanglement scheme and graph diffusion mechanism. The proposed 3D-IDS first employs statistical disentanglement on the traffic features to automatically differentiate tens and hundreds of
complex features and then employs representational disentanglement on the embeddings to highlight attack-specific features. Finally, 3D-IDS fuses the network topology via multi-layer graph diffusion methods for dynamic intrusion detection. Extensive experiments on five benchmarks show the effectiveness of our approach, including binary classifications, multi-classifications, and unknown attack identification. Future work could focus on the deployment of 3D-IDS on future wireless networks.

\section{ACKNOWLEDGEMENT}
This work was supported by National Key R\&D Program of China (Grant No.2022YFB2902200).

\clearpage

\bibliographystyle{ACM-Reference-Format}
\balance
\bibliography{ref}

\clearpage

\appendix
\section{The Input and output }
\label{sec:Input-data-and-ouput-results}
The input instances include statistical features of network traffic, including protocol, flow duration, etc. The output results include confidence scores of the binary labels (benign or abnormal) and multi-class labels (attack labels) for each input. Here we provide an example as follows:
\begin{table}[!htb]
    \centering
    \setlength{\abovecaptionskip}{0cm}
    \caption{An example of the traffic data.}
    \resizebox{\linewidth}{!}{
    \begin{tabular}{ccccccccc}
    \hline
        Src IP & Src Port & Dst IP & Dst Port & Protocol & Flow Duration & ... & Label & Attack \\ \hline
        192.168.1.195 & 65025 & 40.90.190.179 & 443 & 6 & 60155602 & ... & 0 & Benign \\ \hline
    \end{tabular}}
    \label{Input data and output results}
\end{table}

\section{More Details about Eq.5}
\label{sec:Obtainment-of-Eq.5}

Eq.5 in the main paper can be expressed as follows:
$$
	\begin{array}{c}\tilde{\mathbf{w}}=\arg \max \left(\mathbf{w_\textit{N}} \mathcal{F_\textit{N}}-\mathbf{w_\textit{1}} \mathcal{F_\textit{1}}-\sum_{i=2}^{N-1}\left|2 \mathbf{w_\textit{i}} \mathcal{F_\textit{i}}-\mathbf{w_\textit{i+1}} \mathcal{F_\textit{i+1}}-\mathbf{w_\textit{i-1}} \mathcal{F_\textit{i-1}}\right|\right) \\\\
    =\arg \min \left(-L+\sum_{i=2}^{N-1}\left|d_{i+1}-d_{i}\right|\right) \\\\
    =\arg \min \left(\frac{\sum_{i=2}^{N-1}\left|d_{i+1}-d_{i}\right|}{N-1}-\frac{L}{N-1}\right),\end{array}
$$
where $d_i$ refers to the distribution distance of the neighboring features 
$\mathcal{F_\textit{i}}$ and $\mathcal{F_\textit{i+1}}$. Intuitively, Eq.5 aims to maximize the distribution distance between every two features and encourages these features to be evenly distributed in the range for better disentanglement. Therefore, we use Eq.5 as our optimization objective.
Another form of the objective can also be expressed as:
$$\operatorname{argmin}\left(\frac{\sum_{i=1}^{N-1}\left|\mathrm{~d}_{i}\right|}{\mathrm{N}-1}-\frac{\mathrm{L}}{\mathrm{N}-1}\right)\left(5^{*}\right).$$

The difference between the above two expressions lies in the first terms $\frac{\sum_{i=2}^{N-1}\left|d_{i+1}-d_{i}\right|}{N-1}$ and $\frac{\sum_{i=1}^{N-1}\left|d_{i}\right|}{N-1}$. Our Eq.5 in the main paper learns to shape each distance $d_i$ to the optimal distance $\frac{L}{N-1}$, while the latter Eq.5* learns to shape the average distance $\frac{\sum_{i=1}^{N-1}\left|d_{i}\right|}{N-1}$ to the optimal $\frac{L}{N-1}$.

We conduct the following experiment to show the slight performance differences between the above two expressions as Table \ref{two eq}.
\begin{table}[h]
    \caption{F1-score of the two expressions}
    \label{two eq}
    \resizebox{\linewidth}{!}{
    \begin{tabular}{@{}cccc@{}}
    \toprule
     Loss  &EdgeIIoT  &NF-UNSW-NB15-v2 &NF-CSE-CIC-IDS2018-v2 \\ \midrule
     Eq.5  &$96.83$$\pm$\scriptsize$0.36$ &$95.45$$\pm$\scriptsize$0.67$ &$96.34$$\pm$\scriptsize$0.21$ \\
     Eq.5*  &$96.55$$\pm$\scriptsize$0.69$ &$94.22$$\pm$\scriptsize$0.89$ &$95.17$$\pm$\scriptsize$0.68$ \\ \bottomrule
    \end{tabular}}
\end{table}

\section{Disentanglement Studies}
\subsection{Shifting the entangled distribution}
To show that similar feature distributions contribute to lower detection performance, we involved all traffic features for "manual shifting" in the controlled experiments and reported the results in Table \ref{shift}.

To quantify the overlap among features that exist at the beginning and after shifting, we define an average overlap ratio among features. The ratio can be expressed as follows:
$$
\frac{1}{N^{2}} \sum_{i, j}\left(\frac{\text { overlap}_{i,j}}{\operatorname{man}(\text {range}_i, \text {range}_j)}\right), \forall i, j \in[1, N],
$$
where $overlap_{i,j}$ indicates the overlap distance between the min-max normalized feature $i$ and feature $j$. Here the symbol $range_i$ refers to the distribution range of feature $i$, and it can be computed by $[\min (0, \mu-3 \sigma), \max (\mu+3 \sigma, 1)]$, where $\mu$ is the mean and $\sigma$ is the standard deviation. $N$ is the number of features.

Finally, we show that the above ratio on the dataset CIC-ToN-IoT are 62.37\% and 24.16\% respectively before and after "manual shifting", which has a significant reduction after the shifting, i.e., 38.21 points decrease.

\begin{table}[h]
    \caption{F1 score comparisons between methods with raw features and the ones with shift-distribution features, where $*_R$ represents the former and $*_S$ represents the latter. Here LR represents the Logistic Regression method.}
    \centering
    \resizebox{\linewidth}{!}{
    \begin{tabular}{ccccccc}
    \hline
     Attack  &LR\_R  &LR\_S  &E-GraphSage\_R  &E-GraphSage\_S  &3D-IDS\_R &3D-IDS\_S \\ \hline
     Backdoor  &4.21  &8.05  &30.89  &34.46  &52.92	  &54.79  \\
     DoS &53.32  &55.25  &80.76  &81.80  &81.87  &85.53  \\
     Injection  &39.76  &41.76  &86.49  &87.42  &87.39  &87.50  \\
     MITM  &3.82  &6.33  &13.29  &16.08  &49.34  &51.64  \\ \hline
    \end{tabular}}
    \label{shift}
\end{table}

\subsection{Experiments with disentangled features}

The proposed statistical disentanglement can be considered as an explicit automatic feature processing module, aiming to reduce redundant correlations between raw traffic features with more specific guidance.

We also applied such a module to statistical and deep learning models (E-GraphSAGE). The results in Table~\ref{F1 score comparisons} show that our statistical disentanglement also significantly improves attack detection.


\begin{table}[!htb]
    \centering
    \caption{F1 score comparisons between methods with disentangled features (with "+") and with raw features (without "+"), respectively. Here LR represents the Logistic Regression.}
    \resizebox{\linewidth}{!}{
    \begin{tabular}{ccccccc}
    \hline
        Attack & LR & LR+ & E-GraphSage & E-GraphSage+ & 3D-IDS & 3D-IDS+(Ours) \\ \hline
        Backdoor & 4.21 & 13.65 & 30.89 & 35.19 & 52.92 & 64.17 \\ 
        Injection & 39.76 & 50.16 & 86.49 & 87.37 & 87.39 & 87.52 \\ \hline
    \end{tabular}}
    
    \label{F1 score comparisons}
\end{table}

\subsection{The impact of statistical disentanglement}
The proposed feature disentanglement aims to mitigate the redundant correlations between features. It can be considered as "denoising" between features or "shifting" between feature distribution, making the features more distinguishable for better prediction. We show that our 3D-IDS can still capture important patterns jointly contributed by multiple features during the disentanglement.
Here is a more detailed analysis to verify our hypothesis. We conduct experiments on "NF-UNSW-NB15-v2" with an existing pattern that consists of three features including AVG\_THROUGHPUT (traffic throughput), PROTOCOL (protocol), and MAX\_TTL (packet survival time), which we denoted the id of them as A, B, and C, respectively, and the results in Table \ref{importance} show that our 3D-IDS model has a very slight impact on the contribution of such a pattern to attack detection. In experiments, we define an importance score to quantify the impact of the pattern on the detection, and it can be computed by the data entropy variation after using a set of joint features for classification. Importance score can be expressed as:
$$
Importance=Entropy_{sum}-(Entropy_l+Entropy_r),
$$
where the $Entropy_{sum}$ indicates the entropy without joint features, and $Entropy_l+Entropy_r$ refers to the entropy with the joint features.

Table \ref{importance} shows the importance score without and with disentangled features, and we can observe the importance scores of the multiple joined features almost keep the same after disentanglement.

\begin{table}[!htb]
   
    \caption{Features importance score comparisons.}
    \label{importance}
    \resizebox{\linewidth}{!}{
    \begin{tabular}{@{}cccc@{}}
    \toprule
     Feature id  &Features  &Importance score with raw features  &Importance score with disentangled features  \\ \midrule
     1&A+B	  &0.14  &0.13  \\
     2&A+C	  &0.11  &0.11  \\
     3&B+C  &0.09  &0.10  \\
     4&A+B+C  &0.18  &0.18  \\ \bottomrule
    \end{tabular}}
\end{table}

\section{Technical Details}
\subsection{Technical contributions}
This paper has first observed the inconsistent detection performance in existing NIDS and customized a doubly disentangled module to address this issue. Specifically, our key ingredients include a double-feature disentanglement scheme for modeling the general features of various attacks and highlighting the attack-specific features, respectively, and a novel graph diffusion method for better feature aggregation.

\subsection{Ability of unknown attack detection}

The reason 3D-IDS can generalize to unknown attacks is the \textbf{better outlier network traffic (anomaly) detection ability}, which benefits from the proposed double disentanglement and multi-layer graph diffusion modules. The unknown attack detection can be cast as an outlier detection task. A method that can better augment and highlight the deviations in features will better detect the outlier from benign. From this perspective, 3D-IDS first automatically distinguishes hundreds of dimensional features via the proposed statistical entanglement and captures subtle spatiotemporal information via the multi-layer graph diffusion module. Our second disentanglement module guarantees each dimension of features maintains the disentangled properties during the training. By doing so, the attacks with unknown labels and each-dimension-distinguishable features can be properly detected as an outlier, thus improving the ability to generalize to unknown attacks.

\subsection{Incorporating the new nodes}
3D-IDS first appends the new node and the corresponding new edges to the list of nodes and edges. And then we initialize the memory and the representation vector associated with the new node as a $\mathbf{0}$ vector. After the above operations, the node will be incorporated into our methods.

\subsection{Strategies for Data Imbalance}

Data imbalance is an interesting problem to be explored, and we observe such a phenomenon slightly impacts the model performance from experiments. Here are the detailed explanations.
\begin{itemize}
\item Datasets. The public datasets we selected have balanced and imbalanced distributions, such as CIC-ToN-IoT with 47\% Benign samples and EdgeIIoT with 92.72\% Benign samples.
\item Implementation. We use the re-weight trick for the unbalanced-class loss \cite{lin2017focal} in our implementation, which will help improve those unbalanced sample learning. We can also utilize data sampling and data augmentation techniques to alleviate the data imbalance.
\end{itemize}



\section{Handling False Positive Problem}
\label{sec:false-positive-comparsion}
We conduct experiments on the UNSW-NB15 dataset to show the false positives rate. We observe that the false positives rate of our ID-3DS can be as low as 3.27, while the values of the previous methods MLP and E-GraphSAGE are 16.82 and 8.29, respectively. We attribute such a gain to the double disentanglement that can highlight attack-specific features, and thus these more differentiable features help to model the to reduce the false positives.

In addition to model design, the high false positive problem can be seen as a trade-off between false positives and false negatives. We can introduce a learnable parameter to control the tolerance for suspected benign samples according to real-world scenarios.

\section{Real time detection}
\label{sec:real-time-detection}
Our 3D-IDS can be applied to real-time attack detection due to the following reasons.
\begin{itemize}
    \item Training\&Testing. 3D-IDS supports offline model training and online model detection. When the model has been well-trained with training datasets (e.g., the training frequency can be once a week), the trained model can conduct real-time online detection for new observations.
    \item Datasets. As an automatic deep model, 3D-IDS supports different data with features. For real-time anomaly detection in a specific scene, we can use the scene's historical (domain-specific) data as the training set. If the application scenario is similar to our network traffic anomaly detection, the datasets in our paper can be trained for real-time detection.
    \item Efficiency. We conduct a real-time detection on RTX 3090 GPU and report the detection number per minute and accuracy, as shown in Table \ref{realtime}, indicating the potential of 3D-IDS to extend to real-world detection.
    
\end{itemize}

\begin{table}[h]
    \caption{The speed and accuracy of real-time detection in different scenarios.}
    \label{realtime}
    \begin{tabular}{ccc}
    \toprule
     Scenario  &Speed  &Accuracy  \\ \midrule
     Binary Classification	&11334 traffic/min	&0.99  \\
     Multi Classification    &10603 traffic/min	&0.75  \\ \bottomrule
    \end{tabular}
\end{table}

\end{document}